\RequirePackage{ifpdf}
\documentclass[cits]{JINST}
\usepackage{caption}
\usepackage{subcaption}
\usepackage{float}
\usepackage{lineno}
\newcommand{\geant}{{\sc geant4}}
\newcommand{\pythia}{{\sc Pythia}}

\title{Standalone vertex finding in the ATLAS muon spectrometer}

\author{The ATLAS Collaboration}

\abstract{ A dedicated reconstruction algorithm to find decay vertices in the ATLAS muon spectrometer is presented. The algorithm searches the region just upstream of or inside the muon spectrometer volume for multi-particle vertices that originate from the decay of particles with long decay paths.  The performance of the algorithm is evaluated using both a sample of simulated Higgs boson events, in which the Higgs boson decays to long-lived neutral particles that in turn decay to $b\overline{b}$ final states, and $pp$ collision data at $\sqrt{s}$~=~7~TeV collected with the ATLAS detector at the LHC during 2011. }

\begin{document}
\section{Introduction}
This paper describes an algorithm for reconstructing vertices originating from decays of long-lived particles to multiple charged and neutral particles in the ATLAS muon spectrometer (MS)~\cite{ATLASDET}. Such long-lived states are predicted to be produced at the LHC by a number of extensions~\cite{GMSB,RPARITY,IDM,HV1,HV2,HV3} of the Standard Model. The hidden valley scenario~\cite{HV2} is used as a benchmark to study and evaluate the performance of the vertex reconstruction of long-lived particles decaying in the MS. Because of its large volume, the ATLAS MS has good acceptance for a broad range of proper lifetimes and it also has good tracking capabilities at the individual chamber level. A design feature of the MS is low multiple scattering of charged particles, which makes it an ideal instrument for multi-track vertex reconstruction.  The vertex-reconstruction technique described in this paper was a key element in a search for a light Higgs boson decaying to long-lived neutral particles~\cite{HVPRL}.\\
The paper is organized as follows: section 2 describes the muon spectrometer, section 3 discusses the benchmark model and the Monte Carlo (MC) samples, sections 4 and 5 describe the tracking and vertex-finding algorithms and section 6 discusses the performance of the vertex-finding algorithm.

\section{Muon spectrometer}
\label{sec:msgeo}
\begin{figure}[ht!]
  \centering
    \includegraphics[width=125mm]{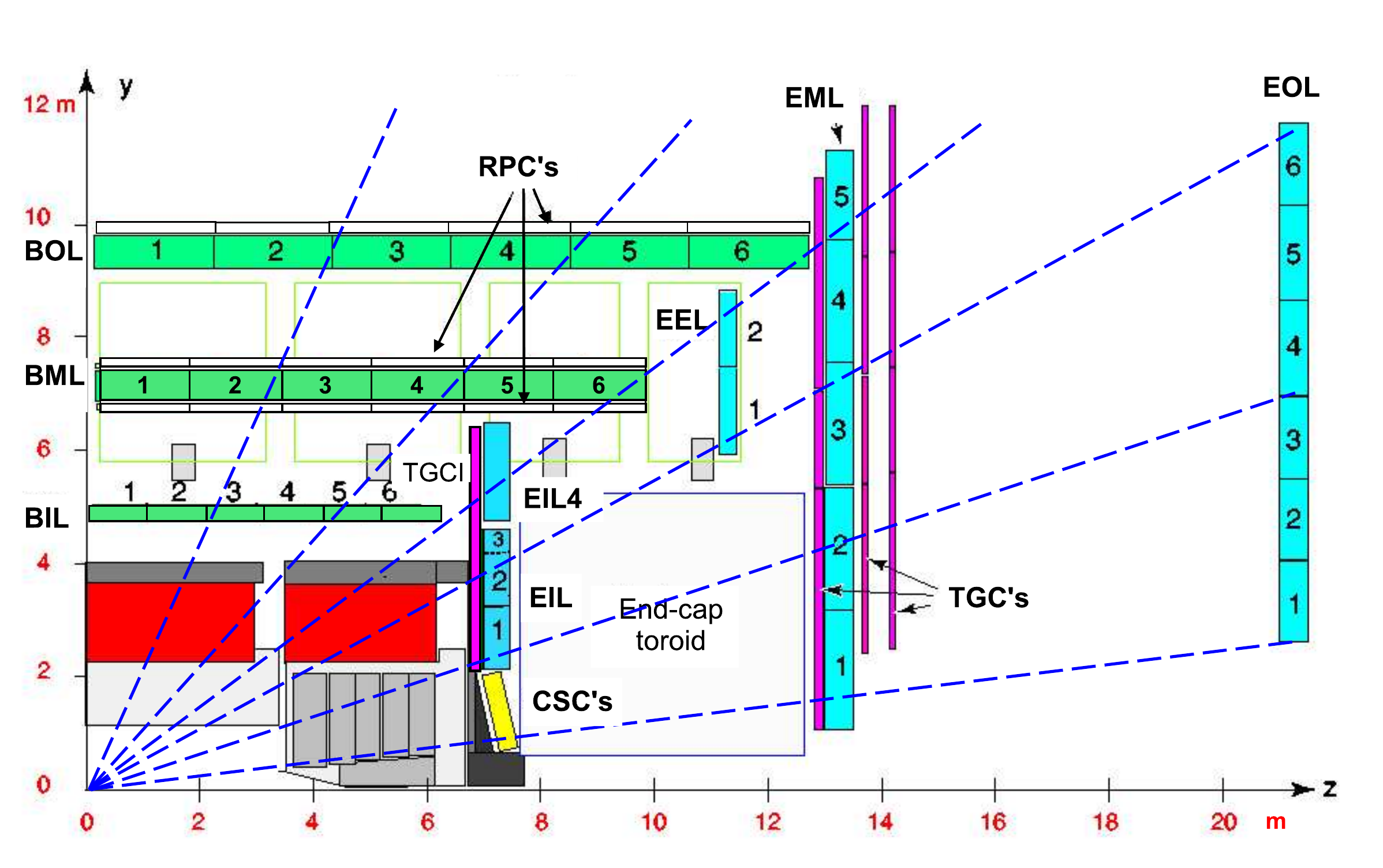}
    \caption{Cross-sectional view of ATLAS in the $r$--$z$ projection at $\phi$~=~$\mathrm{\pi}$/2, from ref.~\cite{ATLASDET}.  The barrel MDT chambers are shown in green, the endcap MDT chambers are blue.  In the barrel (endcaps), the RPC (TGC) chambers are shown outlined in black (solid purple).}
  \label{fig:msprofile}    
\end{figure}

\begin{figure}[ht!]
  \centering
  \includegraphics[width=125mm]{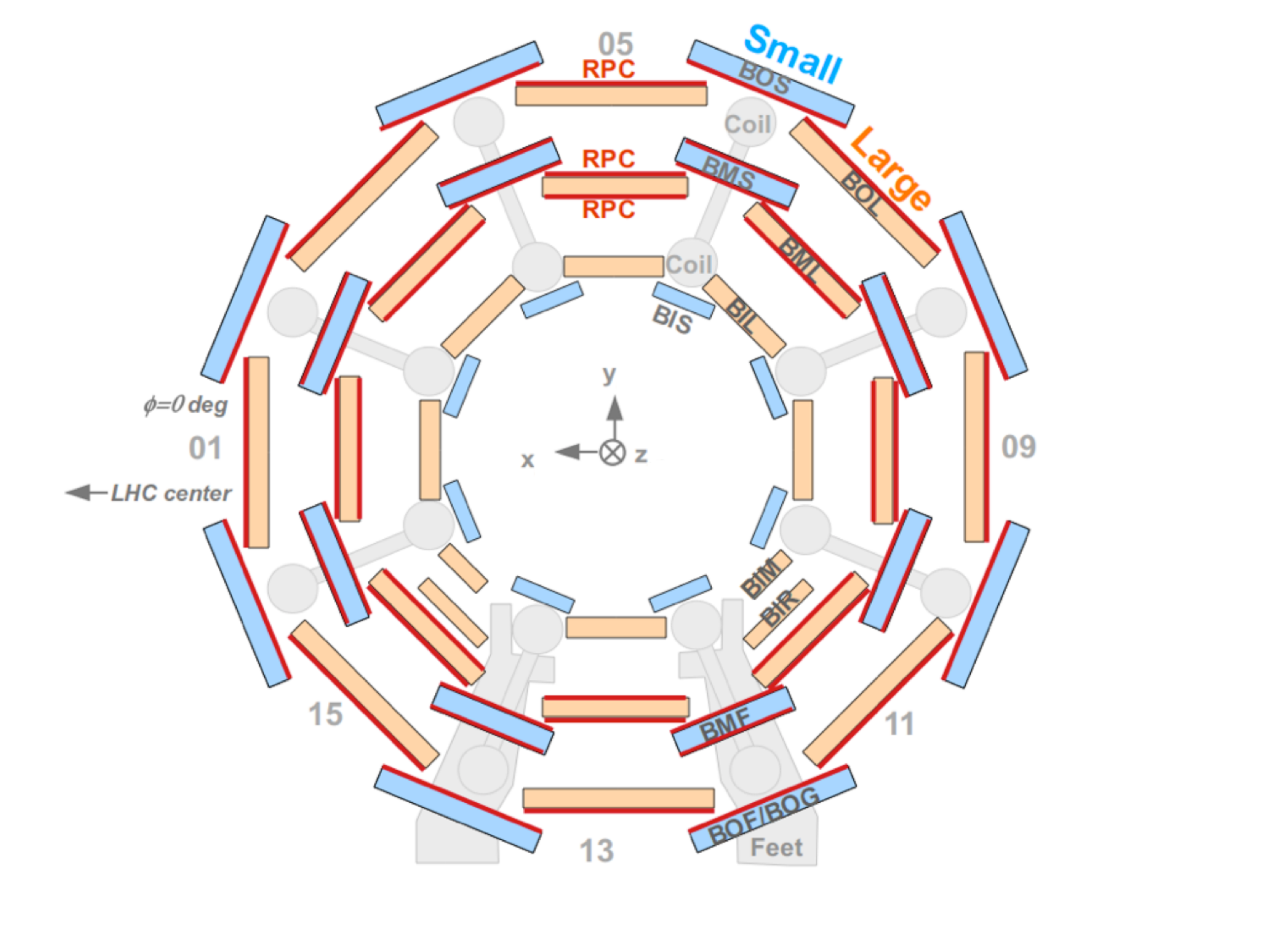}
  \caption{Cross-sectional view of the barrel muon spectrometer perpendicular to the beam axis (non-bending plane).  The MDT chambers in the small sectors are shown in light blue, the MDT chambers in the large sectors are shown in orange and the RPC chambers in red. The eight coils are also visible.}
  \label{fig:mscross}
\end{figure}

ATLAS is a multi-purpose detector~\cite{ATLASDET}, consisting of an inner tracking system (ID), electromagnetic and hadronic calorimeters and a muon spectrometer. The ID is inserted inside a superconducting solenoid, which provides a 2~T magnetic field parallel to the beam direction.\footnote{ ATLAS uses a right-handed coordinate system with its origin at the nominal interaction point (IP) in the centre of the detector and the $z$-axis coinciding with the beam pipe axis. The $x$-axis points from the IP to the centre of the LHC ring, and the $y$-axis points upward. Cylindrical coordinates ($r$,$\phi$) are used in the transverse plane, $\phi$ being the azimuthal angle around the beam pipe. The pseudorapidity is defined in terms of the polar angle $\theta$ as \mbox{$\eta$ = $-$ln tan($\theta$/2)}.}  The electromagnetic and hadronic calorimeters cover the region $|\eta|$~$\leq$~4.9 and have a total combined thickness of 9.7 interaction lengths at $\eta$~=~0. The MS, the outermost part of the detector, is designed to measure the momentum of charged particles escaping the calorimeter in the region \mbox{$|\eta|$~$\leq$~2.7} and provide trigger information for \mbox{$|\eta|$~$\leq$~2.4}.  It consists of one barrel and two endcaps, shown  in figure~\ref{fig:msprofile}, that have fast detectors for triggering and precision chambers for track reconstruction.  Three stations of resistive plate chambers (RPCs) and thin gap chambers (TGCs) are used for triggering in the barrel and endcap MS, respectively.   The precision tracking measurements are provided by monitored drift tube (MDT) chambers throughout the MS and cathode strip chambers (CSCs) in the innermost layer of the endcaps (see figure~\ref{fig:msprofile}).  In the barrel, the precision chambers extend to $|\eta|$~=~1 and are arranged in three cylindrical shells, located at radii of $r$~$\sim$~5.0~m, 7.5~m, and 10.0~m, as shown in figure~\ref{fig:mscross}.  In the endcaps, the precision chambers cover the range 1~$\leq$~$|\eta|$~$\leq$~2.7 and are arranged in three wheels with their faces perpendicular to the $z$-axis located at \mbox{$|z|\sim$7.4~m}, 14.0~m, and 21.5~m.\\
A system of three superconducting toroids (one barrel and two endcap toroids) provides the magnetic field for the MS.  The barrel toroid is 25.0~m in length along the $z$-axis, and extends from $r$~=~4.7~m to 10.0~m in radius; eight superconducting coils  generate the field. The two endcap toroids, each with eight superconducting coils, are inserted in the barrel at each end. They have a length of 5.0 m, an inner bore of 1.65 m and an outer diameter of 10.7 m. The endcap coils are rotated in $\phi$ by 22.5$^{\circ}$ with respect to the barrel coils. Although the MS uses ``air core'' toroid magnets to minimize the amount of material traversed by particles, a non-negligible amount of material is present in the form of support structures, magnet coils and muon chambers.  \\
\subsection{Monitored drift tubes}
\begin{figure}[ht!]
  \centering
  \includegraphics[width=40mm]{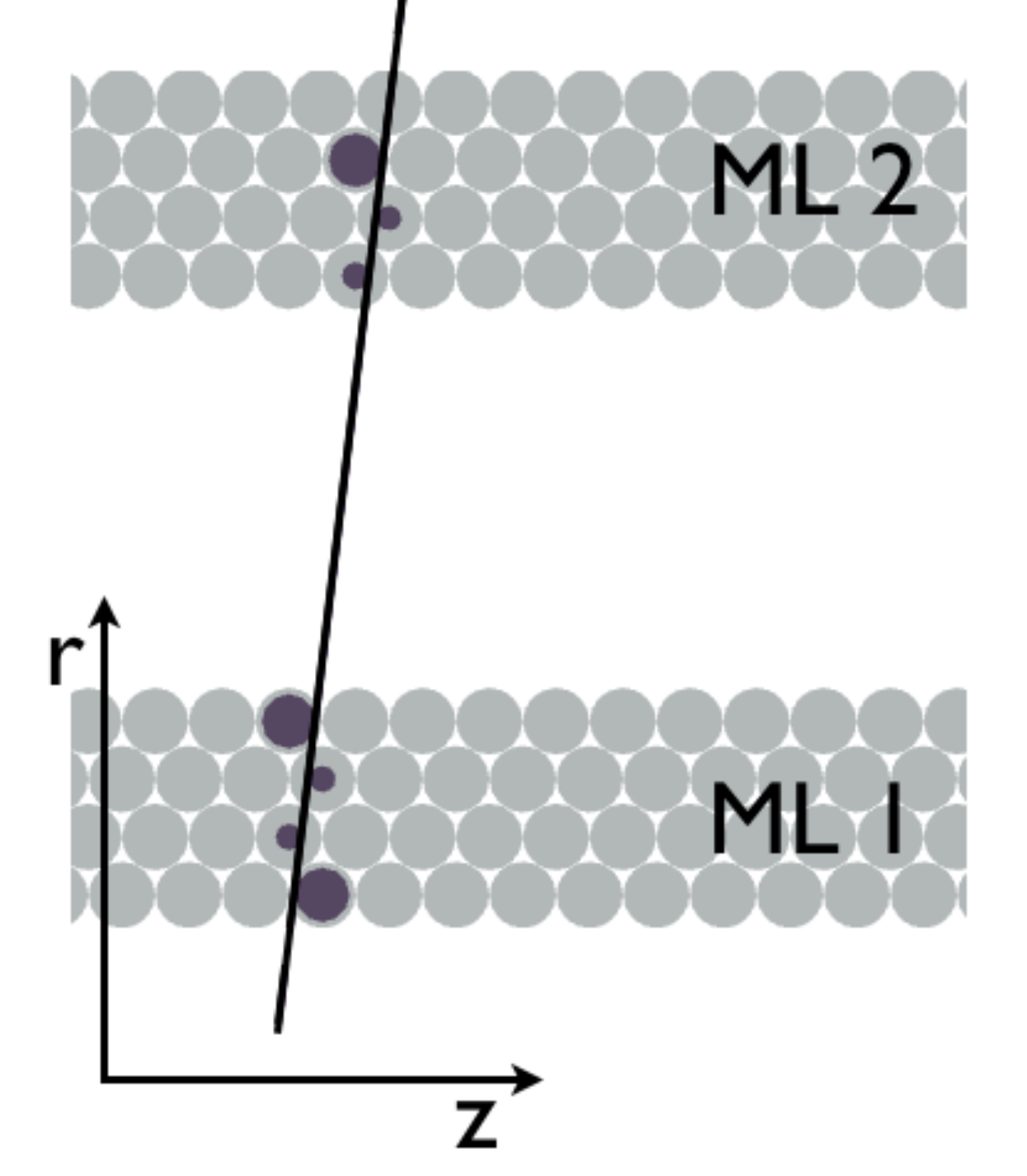}
  \caption{Portion of a muon spectrometer barrel chamber (BIL) with two four-layer multilayers.  The drift circles are shown in dark gray and the charged particle trajectory is shown as the black line.  The spacing between the two multilayers is 170~mm.}
  \label{fig:mdtml}
\end{figure}

In the barrel MS, the MDT chambers are placed around the eight superconducting coils that form the toroid magnet, as shown in figure~\ref{fig:mscross}. In the endcaps, MDT chambers are located either upstream or downstream of the endcap toroids; therefore, all the endcap chambers are located outside the magnetic field region. Because the toroidal magnetic field around the $z$-axis bends trajectories in the $r$--$z$ plane, the MDTs are oriented such that they measure $\eta$ with high precision.\\
The chambers are divided into two types, depending on their position in $\phi$.  Those chambers in the barrel (endcaps) that are located in between the magnet coils are referred to as large (small), while those centred on the magnet coil are small (large). The chamber naming scheme uses a three-letter acronym (e.g. BIL, EMS) to specify a chamber type. The first letter (B or E) refers to barrel or endcap chambers, respectively. The second letter specifies the station (inner, middle or outer) and the third letter refers to the sector (large or small).\\
The MDT chambers consist of two multilayers separated by a distance ranging between 6.5~mm (BIS chambers) and 317~mm (BOS, BOL and BML chambers).  Each multilayer consists of three or four layers of drift tubes.  The individual drift tubes are 30~mm in diameter and have a length of 2--5~m depending on the location of the chamber inside the spectrometer. Each tube is able to measure the drift radius (corresponding to the distance of closest approach of the charged particle to the central wire) with a resolution of approximately 80~$\mu$m~\cite{ATLASDET}.  In each multilayer the charged particle track segment can be reconstructed by finding the line that is tangent to the drift circles. These segments are local measurements of the position and direction of the charged particle.  Figure~\ref{fig:mdtml} shows a charged particle traversing a BIL chamber.\\
Because the tubes are 2--5~m in length, the MDT measurement  provides only a very coarse $\phi$ position of the track hit.  In order to reconstruct the $\phi$ position and direction, the MDT measurements need to be combined with the $\phi$ coordinate measurements from the RPCs (TGCs) in the barrel (endcaps).
\subsection{Trigger chambers}
The RPCs provide the trigger signals and measure the $\phi$ coordinate for segments in the barrel MS.  The chamber planes are located on both sides of the MDT middle stations and one of the two sides of the MDT outer stations. Each chamber has four layers of 3-cm-wide strips, where two layers measure $\eta$ and two layers measure $\phi$, referred to as RPC-eta and RPC-phi planes, respectively. Together, the two planes around the middle stations provide a low transverse momentum ($p_{\rm{T}}$) trigger (up to 10~GeV), and the addition of the chambers in the outer stations allows for high-$p_{\rm{T}}$ triggers.\\
In the endcaps, the trigger signals and $\phi$ measurements are provided by the TGCs.  Each TGC layer consists of cathode strips that measure $\phi$ and anode wires that measure $\eta$.  Measurements from the strips and wires are referred to TGC-phi and TGC-eta measurements, respectively.  The strips have a width of 2--3~mrad, as seen from the interaction point (IP), and the wire-to-wire distance is 1.8~mm.  The middle stations of MDTs in the endcaps are surrounded by seven layers of TGCs, three layers on the IP side and four layers on the opposite side.\\
\subsection{Trigger system}
\label{sec:trigger}
The trigger system~\cite{L1TRIG} has three levels called Level-1 (L1), Level-2 (L2) and the event filter (EF).  The L1 trigger is a hardware-based system using information from the MS trigger chambers, and defines one or more regions-of-interest (RoIs). These are geometrical regions of the detector, identified by ($\eta$,~$\phi$) coordinates, containing potentially interesting physics objects.  The L2 and EF (globally called the high-level trigger, HLT) triggers are software-based systems and can access information from all sub-detectors.  A L1 low-$p_{\rm{T}}$ muon RoI is generated by requiring a coincidence of hits in at least three of the four planes of the two inner RPC planes for the barrel and of the two outer TGC planes for the endcaps. A high-$p_{\rm{T}}$ muon RoI requires additional hits in at least one of the two planes of the outer RPC plane for the barrel and in two of the three planes of the innermost TGC layer for the endcaps.  The muon RoIs have a spatial extent in (\mbox{$\Delta\eta\times\Delta\phi$}) of \mbox{0.2$\times$0.2} in the barrel and \mbox{0.1$\times$0.1} in the endcaps. At the HLT, the L1 RoI information seeds the reconstruction using the precision chamber information, resulting in sharp trigger thresholds up to muon momenta of $p_{\rm{T}}$~=~40~GeV.\\
The Muon RoI Cluster trigger~\cite{TRIGNOTE} is specially designed to select events characterized by a particle decaying to multiple hadrons inside the MS.  This trigger is seeded by a L1 multi-muon trigger that requires at least two muon RoIs with $p_{\rm{T}}$~$\geq$~6~GeV.  At L2, the trigger selects events that have at least three muon RoIs in the barrel clustered in a cone with a radius $\Delta R$~$\equiv\sqrt{(\Delta\eta)^{2}+(\Delta\phi)^2}$~=~0.6.  The Muon RoI Cluster trigger is also required to satisfy track- and jet-isolation criteria~\cite{TRIGNOTE}.  In 2011 data taking, this trigger was only active in the barrel ($|\eta|$~$\leq$~1).

\section{Monte Carlo samples}
\subsection{Benchmark model}
A hidden valley model with a light Higgs boson communicator~\cite{HV2} is used to evaluate the ATLAS detector response to highly displaced decays. In this model, a Higgs boson is produced via gluon fusion and decays to a pair of long-lived neutral pseudoscalars, $H\rightarrow\pi_{\rm{v}}\pi_{\rm{v}}$. Four different MC simulation samples are used for this study, corresponding to different choices of  Higgs boson mass (120~GeV and 140~GeV) and $\pi_{\rm{v}}$ mass (20~GeV and 40~GeV). Because the $\pi_{\rm{v}}$ is a pseudoscalar, it decays  predominantly to heavy fermions, $b\overline{b}$, $c\overline{c}$ and $\tau^{+}\tau^{-}$ in the ratio 85:5:8, as a result of the helicity suppression of the low-mass fermion--antifermion pairs.  The mean proper lifetime of the $\pi_{\rm{v}}$ (expressed throughout this paper as $\tau$ times the speed of light $c$) is a free parameter of the model. In the generated samples, $c\tau$ is chosen so that a sizeable fraction of the decays occur inside the MS.  The \pythia~generator~\cite{PYTHIA} is used to simulate the production and decay of Higgs bosons and the MSTW 2008 leading-order parameterization~\cite{PDF} is used for the parton distribution functions in the protons.  The effect of multiple $pp$ collisions occurring during the same bunch crossing (pile-up) is simulated by superimposing several minimum bias events on the signal event.\footnote{The pile-up was simulated such that the distribution of the number of collision vertices in the MC simulation agrees with the 2011 data.} The generated events are then processed through the full simulation chain based on \geant~\cite{GEANT, ATLSIM}. \\
\subsection{Displaced decays in the MS}
\label{sec:msdecays}
\begin{figure}[ht!]
  \centering
  \includegraphics[width=100mm]{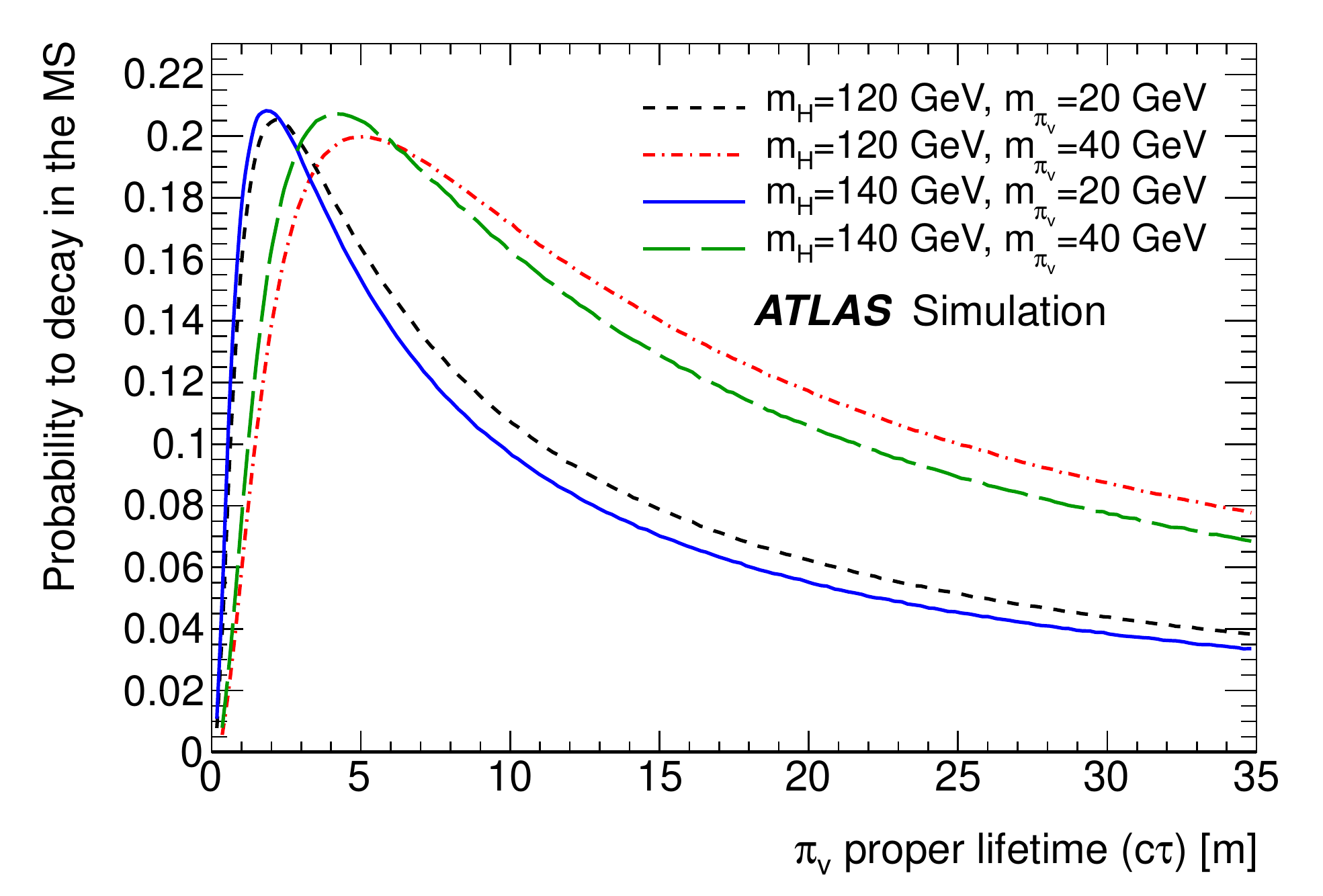}
  \caption{The probability for a $\pi_{\rm{v}}$ to decay inside the fiducial volume of the muon spectrometer as a function of the $\pi_{\rm{v}}$ mean proper lifetime ($c\tau$).}
  \label{fig:decaypos}
\end{figure}

\begin{figure}[h!]
  \centering
  \begin{subfigure}[b]{0.49\textwidth}
    \includegraphics[width=\textwidth]{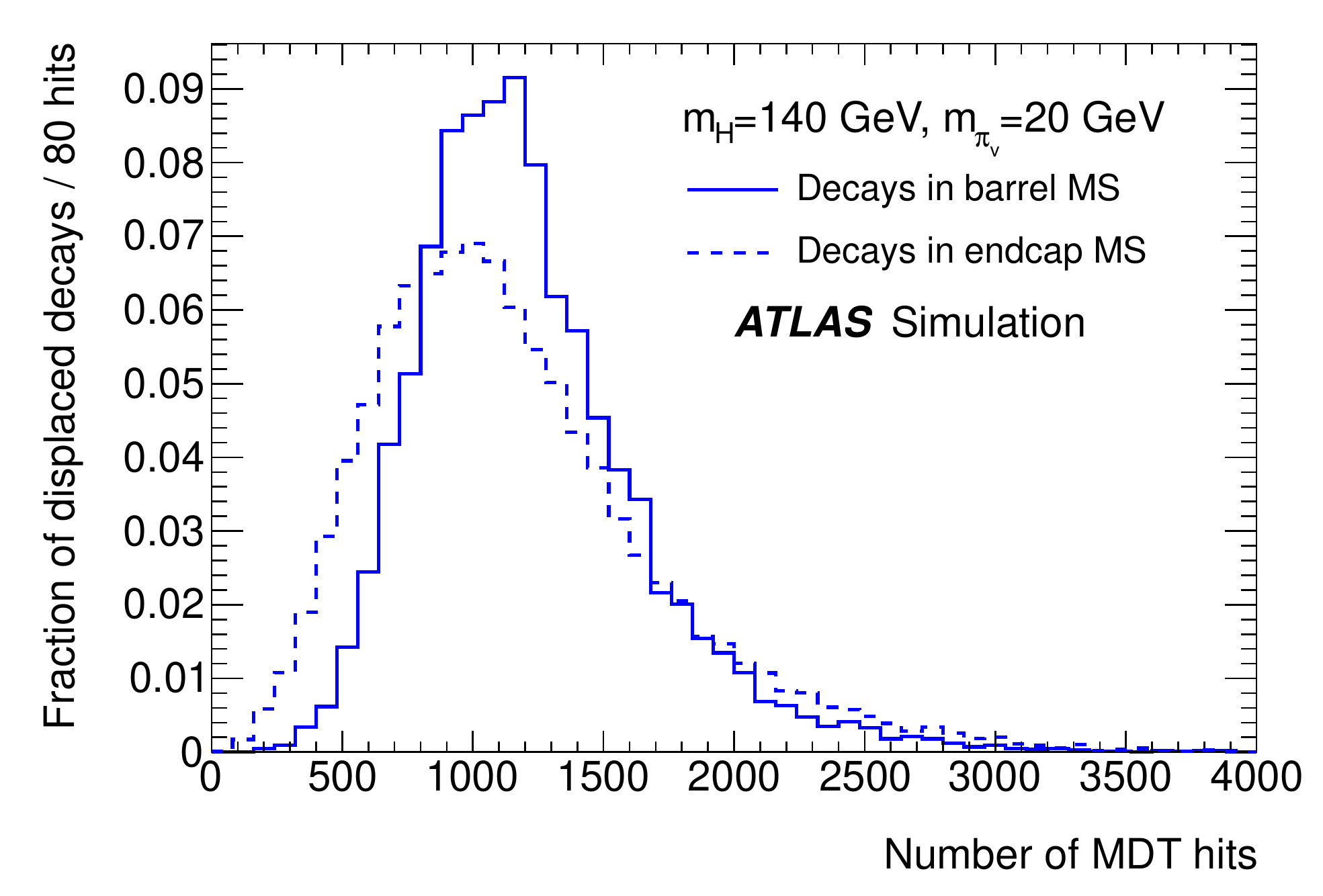}
    \caption{}
  \end{subfigure}
  \begin{subfigure}[b]{0.49\textwidth}
    \includegraphics[width=\textwidth]{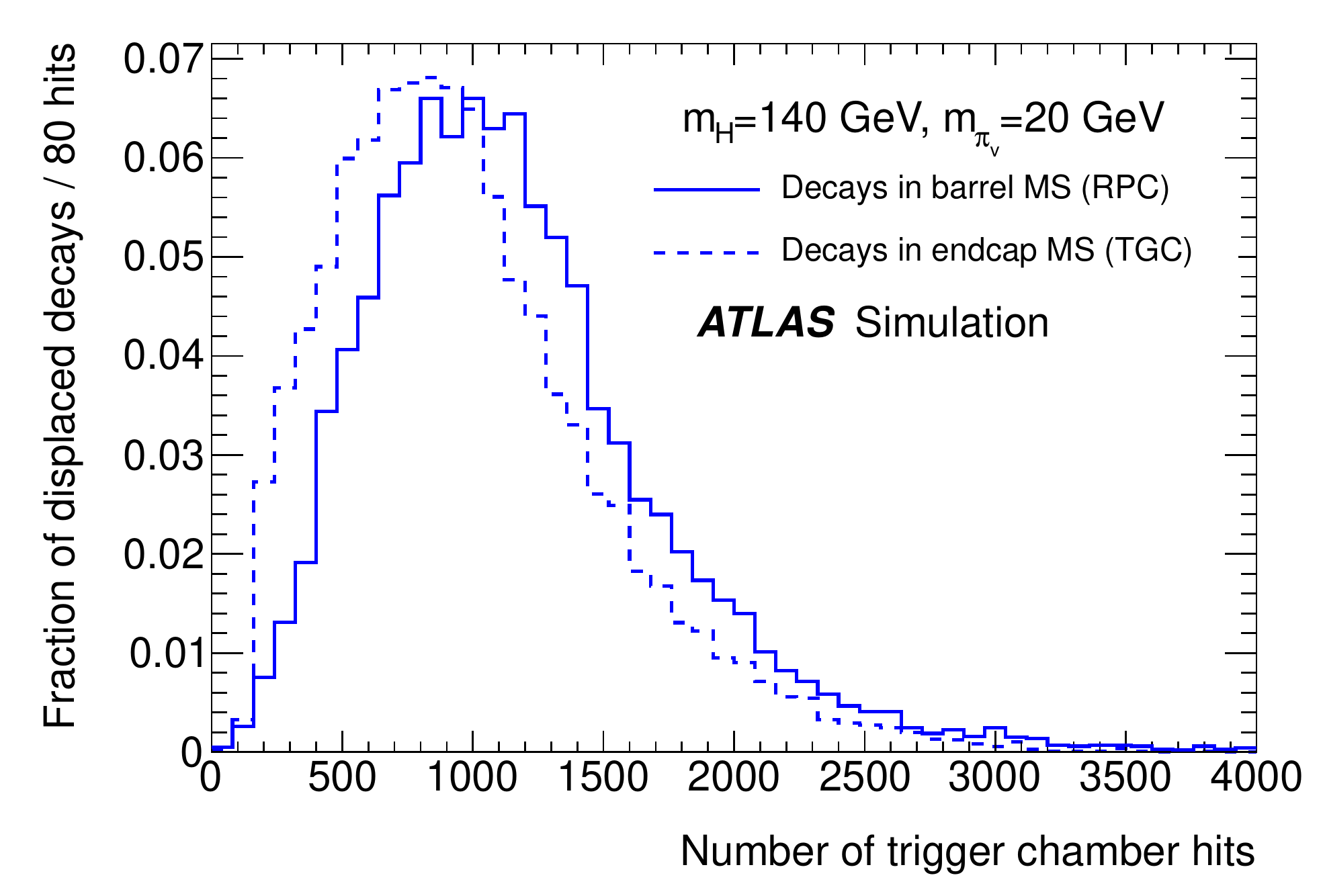}
    \caption{}
  \end{subfigure}
  \caption{Distribution of the total number of (a) MDT and (b) trigger hits per event with a single $\pi_{\rm{v}}$ decay occurring in the barrel ($|\eta|$~$\leq$1) and endcap (1~$<$~$|\eta|$~$<$~2.2) MS. Both plots show the MC mass point $m_{H}$~=~140~GeV, $m_{\pi_{\rm{v}}}$~=~20~GeV.  Similar results are found using the other MC samples.}
  \label{fig:spectrohits}
\end{figure}

For the purposes of triggering on $\pi_{\rm{v}}$ decays in the MS and reconstructing vertices, this study defines the ``MS fiducial volume'' as the region in which $\pi_{\rm{v}}$ decays are detectable.  This fiducial volume is separated into barrel and endcap regions. The barrel MS fiducial volume is defined as the region with $|\eta|$~$\leq$~1, extending from approximately the outermost $\sim$25~cm of the hadronic calorimeter to slightly upstream of the middle station of the MS (3.5~m~$<$~$r$~$<$~7.5~m).  The endcap MS fiducial volume is defined as the region with 1~$<$~$|\eta|$~$<$~2.2, extending from just upstream of the inner endcap muon chambers to the outer edge of the endcap toroids (7~m~$<$~$|z|$~$<$~14~m).  Figure~\ref{fig:decaypos} shows the probability for a $\pi_{\rm{v}}$ to decay inside the MS fiducial volume as a function of the $\pi_{\rm{v}}$ mean proper lifetime ($c\tau$).  This figure indicates that displaced vertices are detectable over a wide range of mean proper lifetimes.
\begin{figure}[ht!]
  \centering
  \includegraphics[width=100mm]{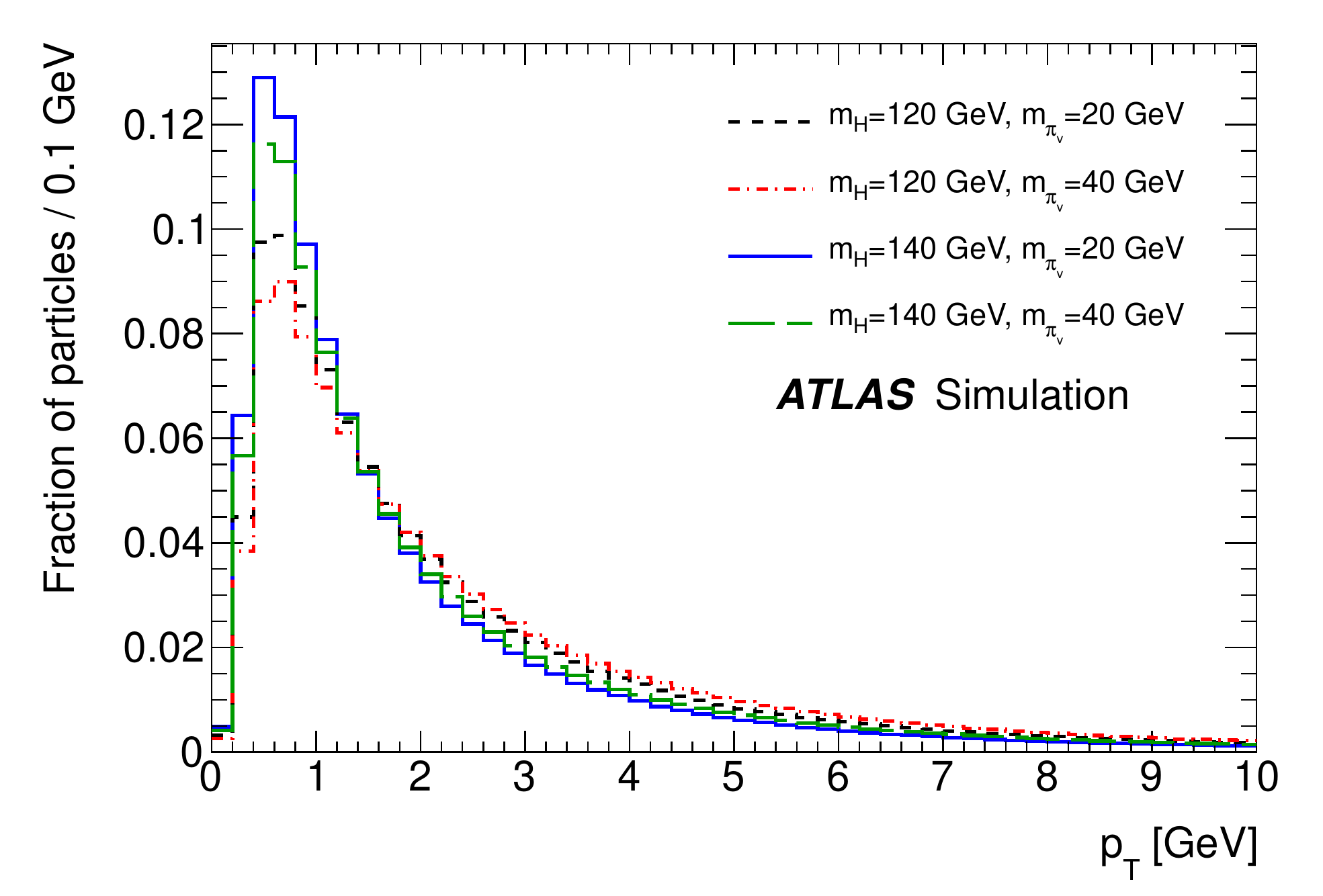}
  \caption{Transverse momentum ($p_{\rm{T}}$) distributions for charged particles from $\pi_{\rm{v}}$ decays, at generator level.}
  \label{fig:mcpt}
\end{figure}%
\\ A decay of a $\pi_{\rm{v}}$ occurring in the MS results in high multiplicity jets of low-$p_{\rm{T}}$ particles (see figure~\ref{fig:mcpt}) produced in a narrow region of the spectrometer.  Typically the decay of a $\pi_{\rm{v}}$ produces a $b\overline{b}$ pair that in turn produces approximately ten charged hadrons and five $\pi^0$ mesons.  Because of these low-$p_{\rm{T}}$ decay products, any decay occurring before the last sampling layer of the hadronic calorimeter would not produce a significant number of tracks in the MS.  Thus, detectable decay vertices are located in the region between the end of the hadronic calorimeter and before the middle station of the muon chambers.  For decays at the end of the hadronic calorimeter, the charged hadrons would traverse, on average, at least two stations (inner and middle); for decays after the inner MDT stations, the charged hadrons would traverse the middle and possibly the outer stations.\\ 
As a consequence of the photons from the $\pi^0$ decays and the non-negligible amount of material in the MS, large electromagnetic (EM) showers are expected to accompany the charged particle tracks from $\pi_{\rm{v}}$ decays in signal events.  The effects of these EM showers can be seen in figure~\ref{fig:spectrohits}, which shows the total number of (a) MDT and (b) RPC (TGC) hits per event with a single $\pi_{\rm{v}}$ decay occurring inside the barrel (endcap) MS.  As these plots show, the MS has an average of $\sim$1000 hits in both the MDT and trigger systems.  The hits are concentrated in a narrow region of the spectrometer, with $\sim$70$\%$ of the hits contained in a cone of radius $\Delta R$~=~0.6 around the $\pi_{\rm{v}}$ line-of-flight.  The average MDT chamber occupancy is approximately 35$\%$ in this cone.
\\A typical muon or $\pi^{\pm}$ traversing the MS leaves a track with 20--25 MDT hits, while the average number of hits per charged particle in displaced decay events is $\sim$100.  This indicates that on average, an event contains $\sim$75$\%$ ``noise'' hits resulting from the EM showers.  These extra hits cause problems for the standard muon-segment-finding routines, which are optimized to find charged tracks in a relatively clean environment.  In order to reconstruct vertices in the MS, efficient tracking, especially at low $p_{\rm{T}}$, is required; therefore a new reconstruction algorithm, capable of reconstructing low-momentum tracks in busy environments, is needed.

\section{Tracklet finding and momentum reconstruction}
\subsection{Tracklet-finding technique}
\label{sec:trackreco}
\begin{figure}[ht!]
  \centering
  \begin{subfigure}[b]{0.49\textwidth}
    \includegraphics[width=\textwidth]{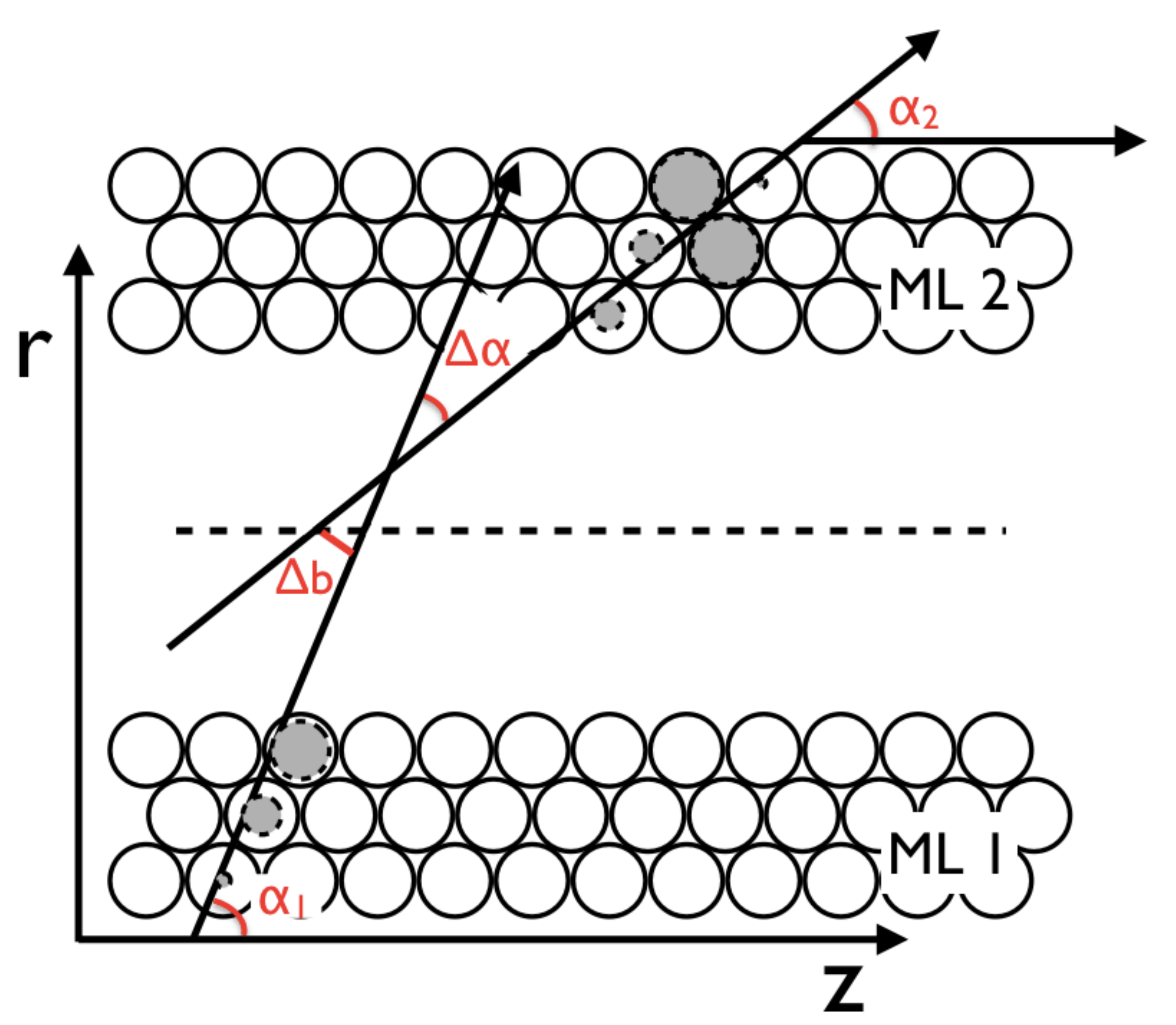}
    \caption{}
  \end{subfigure}
  \begin{subfigure}[b]{0.49\textwidth}
    \includegraphics[width=\textwidth]{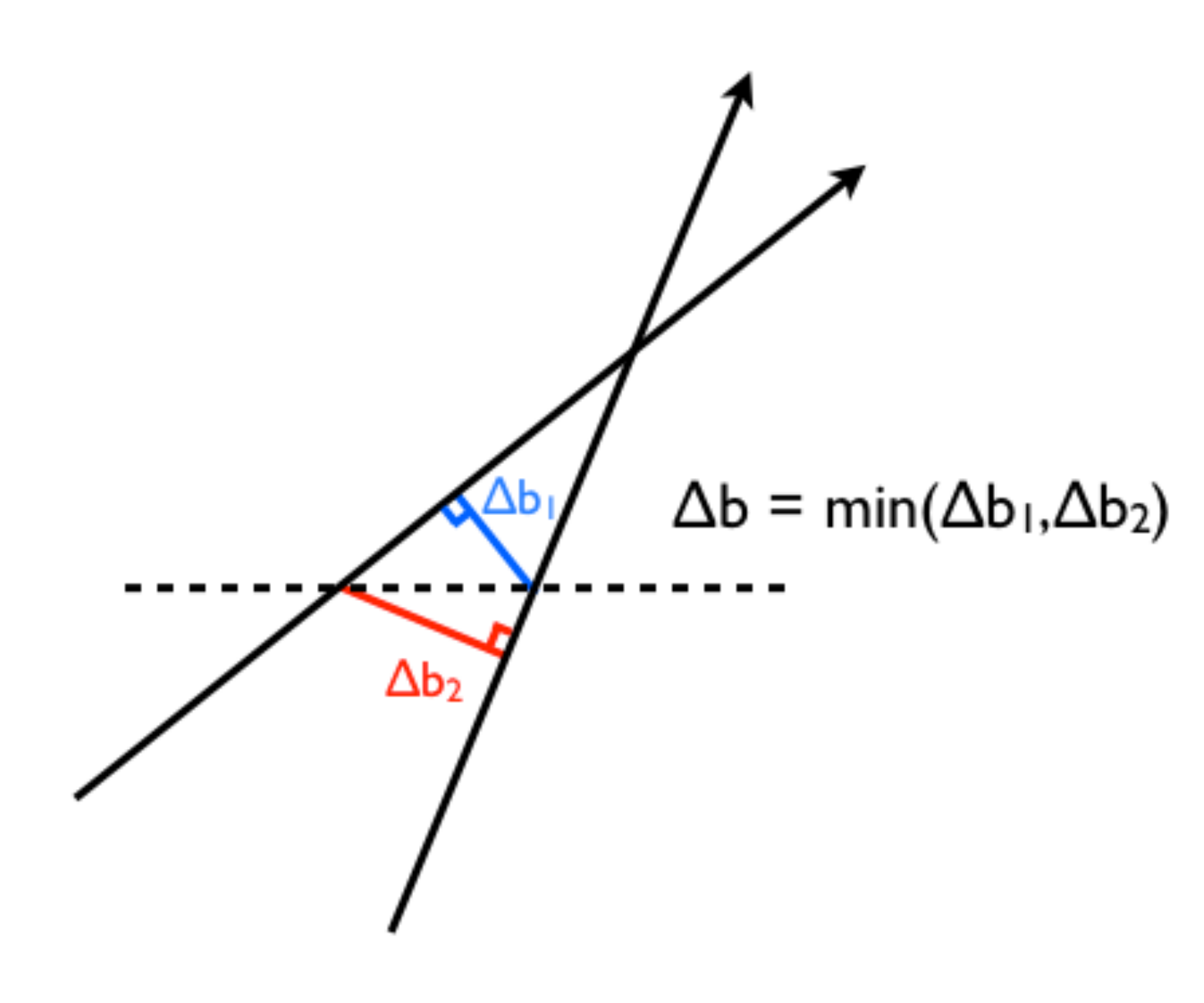}
    \caption{}
  \end{subfigure}
  \caption{Schematic of a MS barrel chamber with one segment in multilayer (ML) 1 and one in multilayer 2.  Shown in (a), the two single-multilayer segments reconstructed in the respective multilayers and (b) a close-up of the middle plane of the chamber. The variable $\alpha_{1(2)}$ is defined as the angle with respect to the $z$-axis of the segment in multilayer 1 (2).  The parameter $\Delta\alpha$ is defined as $\Delta\alpha$~$\equiv$~$\alpha_{1}$~$-$~$\alpha_{2}$ and $\Delta b$ is defined to be the distance of closest approach between the pair of segments at the middle plane of the MDT chamber.  The middle plane of the chamber is the plane equidistant from multilayers 1 and 2, represented here by the dashed line.}
  \label{fig:delta}
\end{figure}

The spatial separation between the two multilayers inside a single MDT chamber provides a powerful tool for pattern recognition.  The specialized tracking algorithm presented here exploits this separation by matching segments from multilayer 1 (ML1) with those from multilayer 2 (ML2).  The algorithm starts by reconstructing single-multilayer straight-line segments that contain three or more MDT hits using a minimum $\chi^2$ fit.  All segments that have a $\chi^2$ probability greater than 5$\%$ are kept.\\ 
In order to pair the segments belonging to the same charged particle, segments from ML1 are matched with those from ML2 using two parameters, $\Delta b$ and $\Delta\alpha$, as shown in figure~\ref{fig:delta}(a).  The parameter $\Delta b$ is taken to be the minimum of the two possible distances between the point where one segment crosses the middle plane and the line defined by the other segment, as illustrated in figure~\ref{fig:delta} (b).\footnote{For the purposes of plotting, the parameter $\Delta b$ is signed according to a local coordinate system with an origin defined as the point where the ML1 segment crosses the middle plane.  In all other cases, $\Delta b$ is a positive definite quantity.}  For chambers in the magnetic field, $\Delta b$~$\sim$~0 selects segments that are tangent to the same circle and hence belong to the same charged particle.
 The second parameter, $\Delta\alpha$~$\equiv$~$\alpha_{1}$~$-$~$\alpha_{2}$, is the angle between the two segments.  It can only be used for matching segments in the case of chambers outside the magnetic field region. If the chamber is inside the magnetic field region, $\Delta\alpha$ is the bending angle of the track inside the chamber and can be used to measure the track momentum for low-$p_{\rm{T}}$ particles (see section~\ref{sec:momentum}).  In the following, this paired set of single-multilayer segments and corresponding track parameters is referred to as a \emph{tracklet}.  Because of the large number of RPC (TGC) hits in signal events, the RPC-phi (TGC-phi) hits cannot be associated with the MDT barrel (endcap) tracklets. Consequently the tracklets reconstructed using this method do not have a precise $\phi$ coordinate or direction. Therefore, the tracklets are assigned the $\phi$ coordinate of the MDT chamber centre and are assumed to be travelling radially outward.\\
The intrinsic angular resolution of the single-multilayer MDT segments is derived from a fully simulated MC sample of high-momentum charged particles\footnote{This sample contains single muons with $p$~=~1~TeV, distributed evenly in the ranges $-\pi$~$<$~$\phi$~$<$~$\pi$ and $|\eta|$~$<$~1 and is simulated using the full ATLAS simulation chain.} that produce straight-line segments in the MDT chambers.  From this sample, the parameters $\Delta\alpha$ and $\Delta b$ are determined to have RMS values of 4.3~mrad and 1.0 mm, respectively, for segments containing three MDT hits.\\
The tracklet selection criteria are listed in table~\ref{tab:trackletcuts} for each of the muon chamber types. The variable $\Delta\alpha_{\rm{max}}$ refers to the maximum amount of bending that is allowed inside the chamber for the two single-ML segments to be considered matched and corresponds to a minimum tracklet momentum of 0.8~GeV for chambers that are located inside the magnetic field.  Tracklets are refit as a single straight-line segment spanning both multilayers if their $|\Delta\alpha|$ is less than 12~mrad.\\
\begin{table}[ht!]
  \centering
  \begin{tabular}{c|c|c|c|c|c}
    Chamber & Number of & ML Spacing & $|\Delta\alpha_{\rm{max}}|$ & $|\Delta b$$_{\rm{max}}|$ & Refit \\
    Type & Layers & (mm) & (mrad) & (mm) & \\\hline\hline
    BIS & 4 & 6.5 & 12 & 3 & Always \\\hline
    BIL & 4 & 170  & 36 & 3 & if $|\Delta\alpha|$~$<$~12~mrad \\\hline
    BMS & 3 & 170  & 67 & 3 & if $|\Delta\alpha|$~$<$~12~mrad \\\hline
    BML & 3 & 317 & 79 & 3 & if $|\Delta\alpha|$~$<$~12~mrad \\\hline
    BOS & 3 & 317 & 12 & 3 & Always \\\hline
    BOL & 3 & 317 & 36 & 3 & if $|\Delta\alpha|$~$<$~12~mrad \\\hline
    Endcap & 3 & 170 & 12 & 3 & Always\\\hline
  \end{tabular}
  \caption{Relevant chamber parameters and the selection criteria for reconstructing tracklets in each of the MDT chamber types.  Tracklets are refit as a single straight-line segment spanning both multilayers if they satisfy the criterion listed in the ``Refit'' column.}
  \label{tab:trackletcuts}
\end{table}
In the endcaps, and in the BIS and BOS chambers in the barrel MS, the MDT chambers are outside the magnetic field region; therefore, segment pairs from these chambers are combined and refit as a single straight-line segment, containing at least six MDT hits.   The combined segments result in a 0.2~mrad angular resolution compared to the 4.3~mrad resolution obtained with the two single-multilayer segments. This improvement in the angular resolution is due to the increased lever arm and additional MDT hits when fitting the segments spanning the two MLs.  
\subsection{Momentum and charge measurements}
\label{sec:momentum}
For segments found in the MS barrel chambers in the magnetic field, the measurement of $\Delta\alpha$ can be used to calculate the tracklet momentum.  The tracklet momentum can be determined using a relation of the form $p$~=~$k$/$|\Delta\alpha|$, where the parameter $k$ is derived for each muon chamber type.  From the uncertainty in $\Delta\alpha$, calculated for each segment pair as $\sigma_{\rm{\Delta\alpha}}$~$\equiv$~$\sqrt{(\sigma_{\alpha_{1}})^2 + (\sigma_{\alpha_{2}})^2}$, the uncertainty in the momentum measurement can be shown to be $\sigma_p$/$p$~$\approx$~0.06$\cdot$$p$/GeV in the BML chambers, $\sigma_p$/$p$~$\approx$~0.08$\cdot$$p$/GeV in the BMS chambers and $\sigma_p$/$p$~$\approx$~0.13$\cdot$$p$/GeV in the BOL and BIL chambers.\\
The tracklet charge is obtained from  \mbox{$q$~=~sign($\Delta\alpha\cdot$$z$$\cdot$$\hat{p_{z}}$)}, where $\Delta\alpha$ is the bending angle, $z$ is the position of the tracklet and $\hat{p_{z}}$ is the direction of the tracklet as measured in ML1.  The charge of the particle can be identified with an efficiency greater than 90$\%$ for reconstructed tracklets with momentum less than 7~GeV in the BML chambers, 5~GeV in the BMS chambers and 3~GeV in the BOL and BIL chambers.\\
\subsection{Application of the tracking algorithm in MC signal events}
\begin{figure}[ht!]
  \centering
  \begin{subfigure}[b]{0.47\textwidth}
    \includegraphics[width=\textwidth]{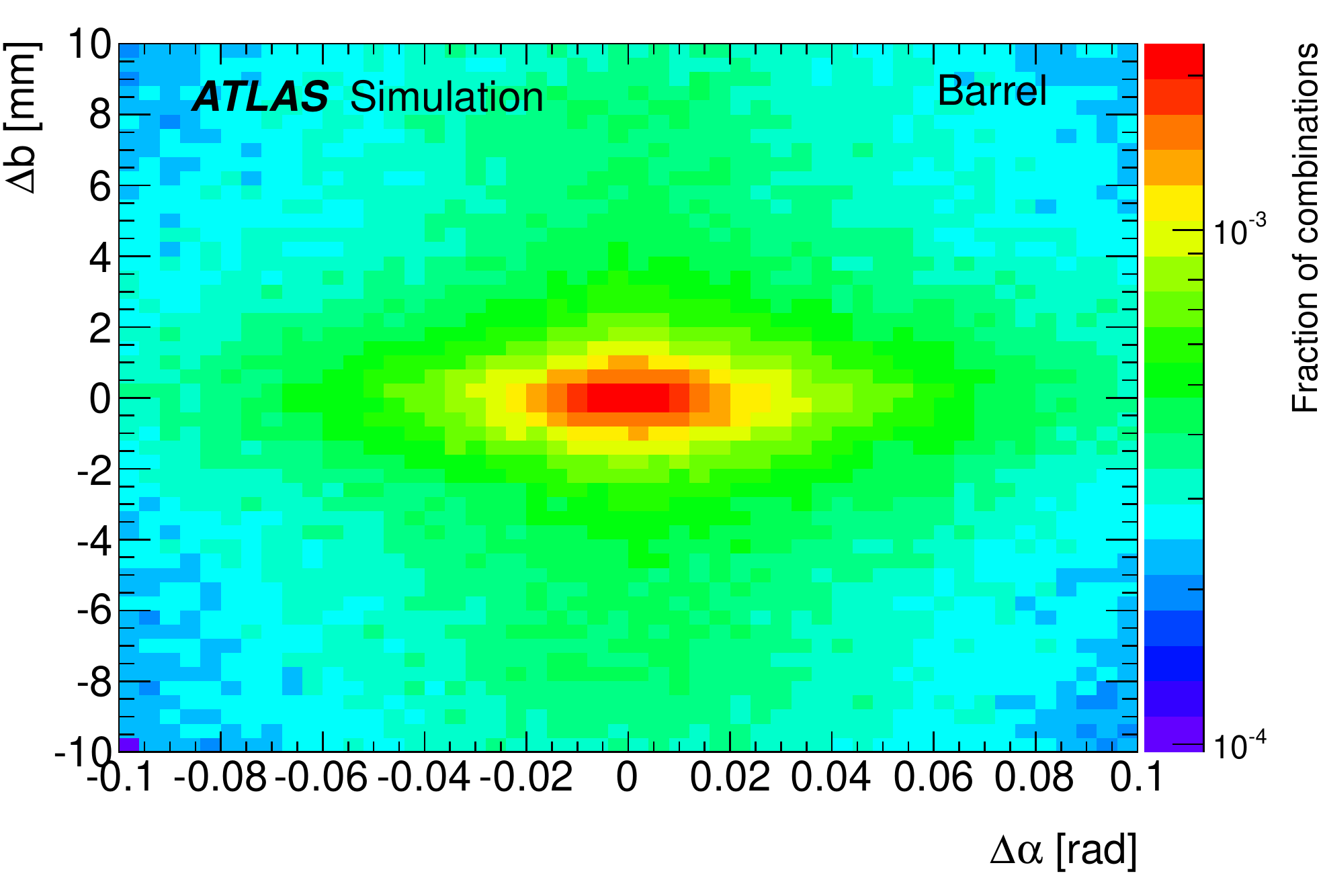}
    \caption{}
  \end{subfigure}
  \begin{subfigure}[b]{0.47\textwidth}
    \includegraphics[width=\textwidth]{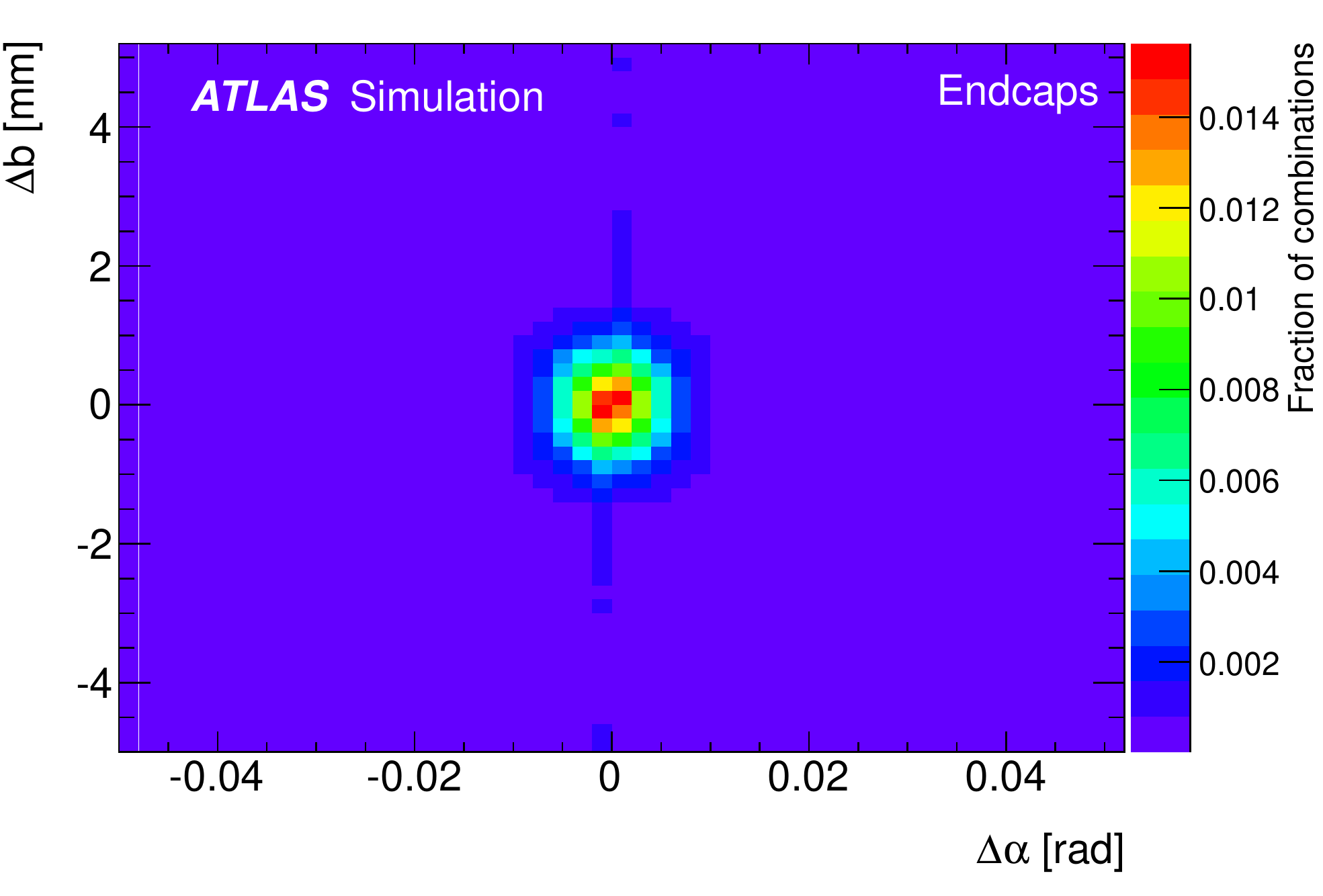}
    \caption{}
  \end{subfigure}
  \caption{Distributions of $\Delta b$ vs $\Delta\alpha$ from MC signal events for segment combinations (a) in the barrel MS region ($|\eta|$~$<$~1) and (b) in the endcap region (1~$<$~$|\eta|$~$<$~2.7). Both plots are normalized to unit area and show the MC mass point $m_{H}$~=~140~GeV, $m_{\pi_{\rm{v}}}$~=~20~GeV.  Similar results are found using the other MC samples.}
  \label{fig:signal_da_db}
\end{figure}

\begin{figure}[ht!]
  \centering
  \includegraphics[width=100mm]{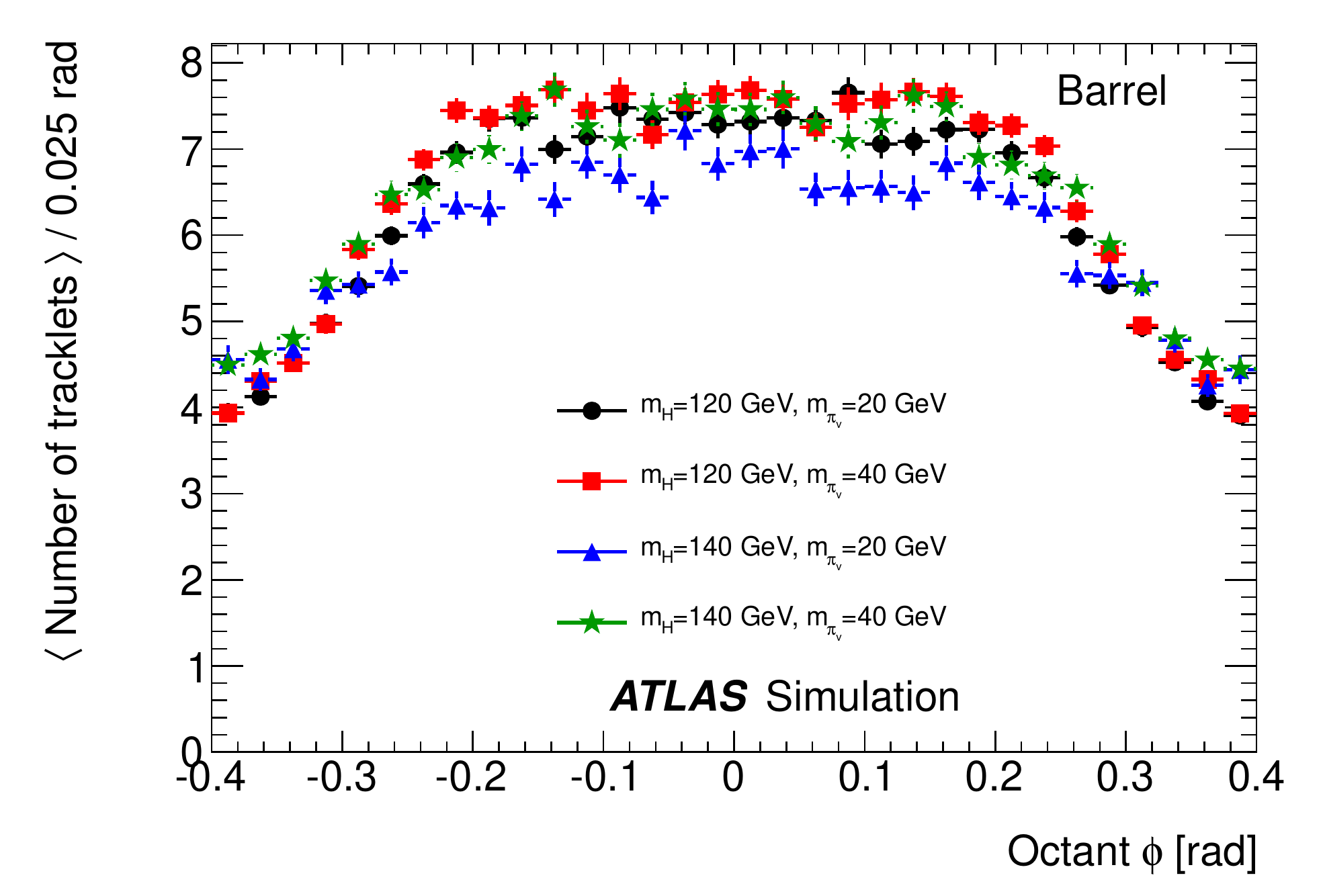}
  \caption{The average number of tracklets with a charge and momentum measurement reconstructed as a function of the position, in $\phi$, of the displaced decay, for decays occurring in the barrel MS. All sectors have been projected such that the centre of a large sector corresponds to $\phi$~=~0 and the centre of a small sector corresponds to $\phi$~=~$\pm\mathrm{\pi}$/8.}
  \label{fig:ntrackletphi}
\end{figure}

\begin{figure}[ht!]
  \centering
  \begin{subfigure}[b]{0.49\textwidth}
    \includegraphics[width=\textwidth]{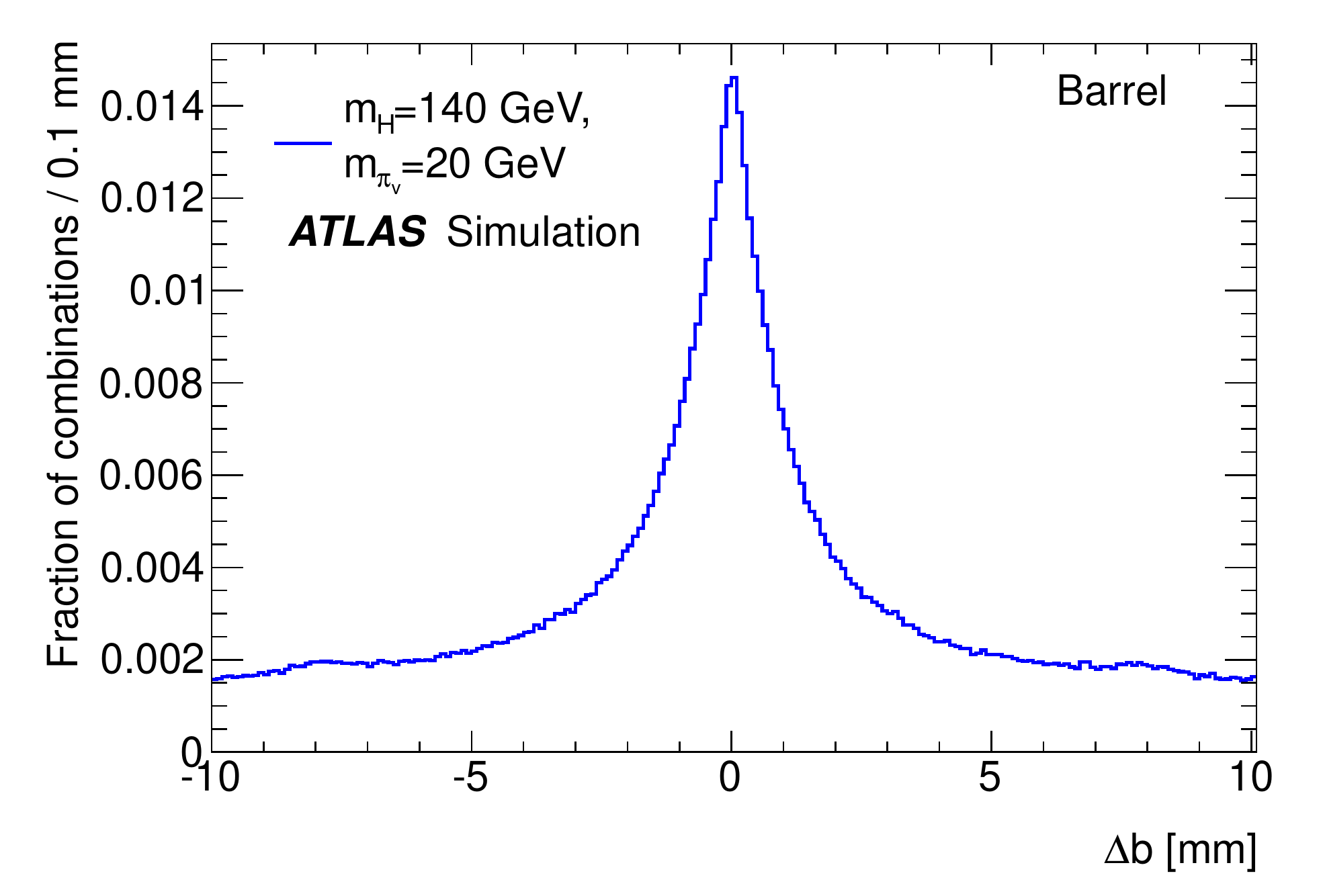}
    \caption{}
  \end{subfigure}
  \begin{subfigure}[b]{0.49\textwidth}
    \includegraphics[width=\textwidth]{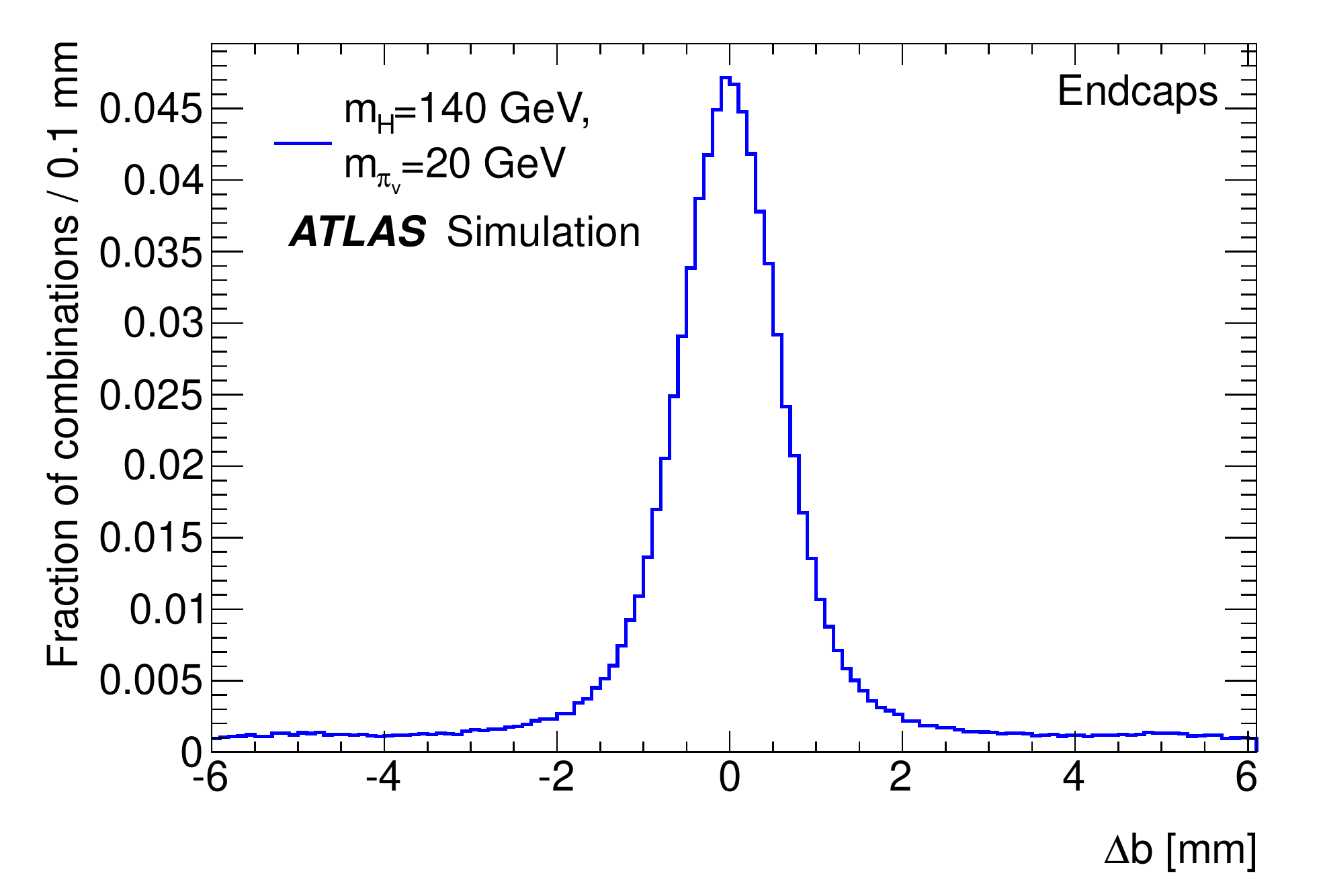}
    \caption{}
  \end{subfigure}
  \caption{Distributions of $\Delta b$ for segment combinations in (a) the barrel MS region ($|\eta|$~$<$~1) and (b) the endcap MS region (1~$<$~$|\eta|$~$<$~2.7) for MC signal events.  Segment combinations are required to satisfy the appropriate $\Delta\alpha$ criterion listed in table~\protect\ref{tab:trackletcuts}.  In the case of the endcap combinations, they are required to have been successfully refit to a single straight-line with $\chi^2$ probability greater than 5$\%$. Both plots show the MC mass point $m_{H}$~=~140~GeV, $m_{\pi_{\rm{v}}}$~=~20~GeV.  Similar results are found using the other MC samples.}
  \label{fig:signaldeltab}
\end{figure}

The performance of the tracklet reconstruction algorithm has been be studied using the MC signal events that have a displaced decay occurring inside the fiducial volume of the MS.  Figure~\ref{fig:signal_da_db} shows the two-dimensional distribution of $\Delta b$ versus $\Delta\alpha$ for tracklets in the barrel and endcap regions.  The segment combinations corresponding to real charged particles can be seen in the central region, while the diffuse background comes from the incorrect pairing of segments.  This reconstruction method finds an average of nine (eight) tracklets per displaced decay in the barrel (endcaps) fiducial volume.
 In the barrel MS an average of six tracklets have an associated momentum measurement. Figure~\ref{fig:ntrackletphi} shows the average number of tracklets with a momentum measurement reconstructed as a function of the displaced decay position, in $\phi$, for one octant of the barrel MS.  The extra material present in the barrel small sectors creates large EM showers that affect the reconstruction algorithm.  As a result, near the centre of the small sectors ($\phi$~$\sim$~$\pm$0.4) there are approximately half as many tracklets reconstructed, on average, compared to a displaced decay occurring near the centre of the large sectors ($\phi$~$\sim$~0).  In contrast, decays occurring in the fiducial volume of the endcap MS produce an average of eight tracklets, without momentum measurements, independent of the $\phi$ coordinate of the decay.\\
Figure~\ref{fig:signaldeltab}(a) shows the distribution of $\Delta b$ in the barrel MS region for all segment combinations that satisfy the criteria for $|\Delta\alpha|$ listed in table~\ref{tab:trackletcuts}.  Figure~\ref{fig:signaldeltab}(b) shows the distribution of $\Delta b$ in the endcap region for all segment combinations that have $|\Delta\alpha|$~$<$~12~mrad and have been successfully refit as a single straight-line segment with $\chi^2$ probability greater than 5$\%$.  The fraction of fake tracklets reconstructed can be estimated by using the side bands to measure the combinatorial background.  The side bands are fit to a straight-line that is extrapolated to the signal region.  Taking the ratio of the number of tracklets under the background fit to the total number of tracklets in the signal region gives a fake rate of $\sim$25$\%$ in the barrel and $\sim$5$\%$ in the endcaps.  The lower fake rate in the endcap is due to the refit procedure, which selects only those combinations of single-ML segments that can be fit as a single straight-line segment with $\chi^2$ probability greater than 5$\%$.

\section{Vertex reconstruction}
\label{sec:vxfinding}
Most of the MS barrel chambers ($|\eta|$~$<$~1) are immersed in a magnetic field while the endcap chambers (1~$<$~$|\eta|$~$<$~2.7) are all outside the field region.  Thus, all tracklets reconstructed in the endcaps have no associated momentum measurement, while in the barrel, most of the tracklets reconstructed from two segments have a $\Delta\alpha$ measurement that provides an estimate of the momentum.  Due to this difference in tracklet reconstruction, different vertex-reconstruction algorithms are employed in the MS barrel and endcaps.  In both cases, the algorithms have been tuned to maximize the vertex-finding efficiency at the expense of vertex-position resolution.\\
The barrel and endcap vertex-reconstruction algorithms, described in detail in the following sections, proceed as follows:
\begin{enumerate}
  \item Tracklets are reconstructed in individual chambers.
  \item The tracklets from all chambers are grouped into clusters using a cone algorithm.
  \item The lines-of-flight in $\eta$ and $\phi$ of the long-lived particle are calculated using the tracklets and RPC-phi (barrel) or TGC-phi (endcaps) hits, respectively.
  \item The clustered tracklets are mapped onto a single $r$--$z$ plane as defined by the $\phi$ line-of-flight.
  \item The vertex position is reconstructed by back-extrapolating the tracklets; in the barrel the tracklets are extrapolated through the magnetic field, while in the endcaps the tracklets are extrapolated as straight lines.
\end{enumerate}
The barrel and endcap vertex-reconstruction algorithms exclusively use the tracklets reconstructed in the barrel and endcap, respectively.  For displaced decays occurring near $|\eta|$~=~1, both algorithms can independently reconstruct vertices.  In case there are two reconstructed vertices, it is left to the analysis make the final vertex selection. 
\subsection {Vertex reconstruction in the barrel MS}
\label{sec:barvxreco}
\begin{figure}[ht!]
  \centering
  \begin{subfigure}[b]{0.49\textwidth}
    \includegraphics[width=\textwidth]{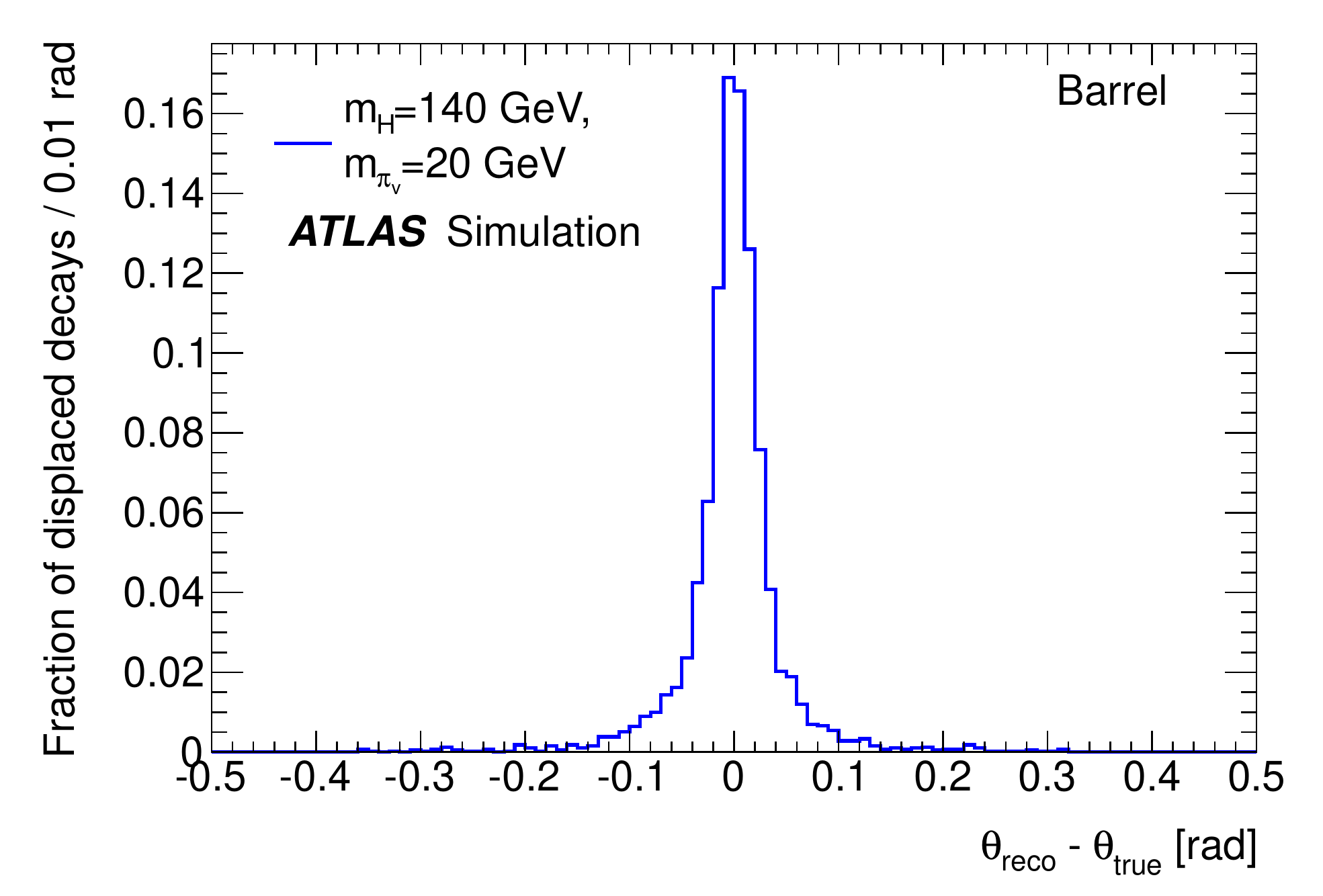}
    \caption{}
  \end{subfigure}
  \begin{subfigure}[b]{0.49\textwidth}
    \includegraphics[width=\textwidth]{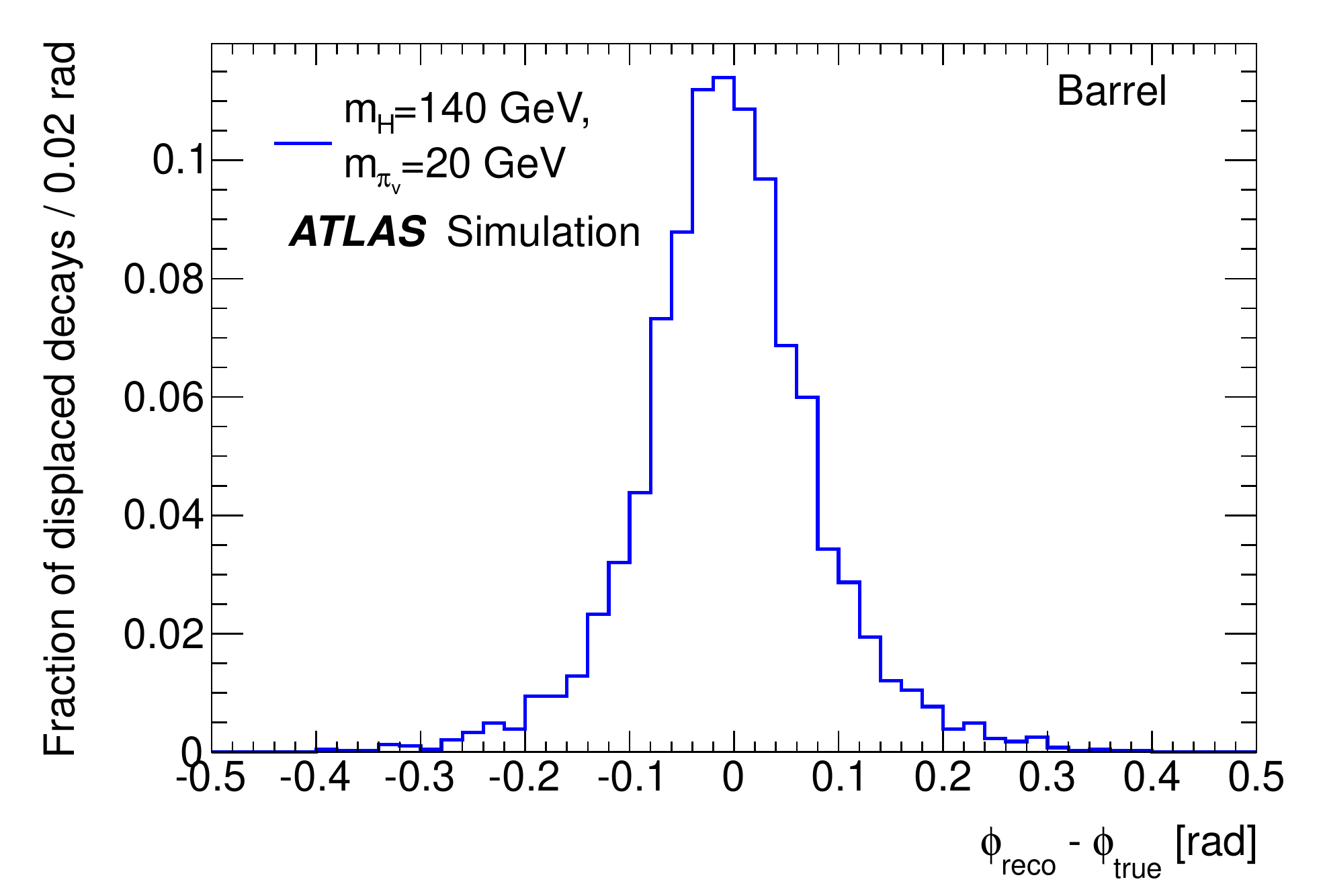}
    \caption{}
  \end{subfigure}
  \caption{Distributions of the flight direction residuals in the (a) $\theta$ and (b) $\phi$ directions. Both plots show the MC mass point $m_{H}$~=~140~GeV, $m_{\pi_{\rm{v}}}$~=~20~GeV.  Similar results are found using the other MC samples.}
  \label{fig:lineOfFlight}
\end{figure}

\begin{figure}[ht!]
  \centering
  \includegraphics[width=125mm]{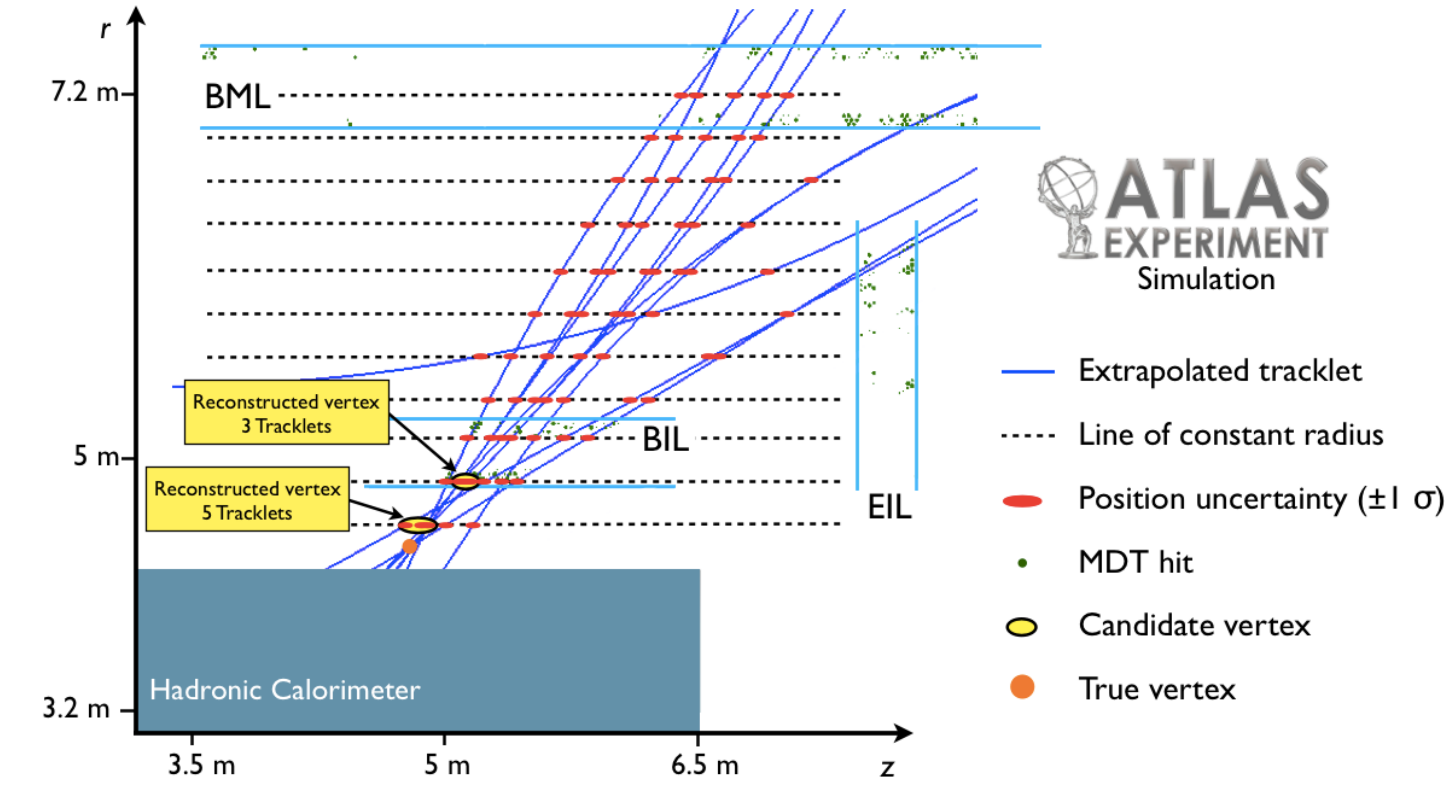}
  \caption{Diagram illustrating the reconstruction technique employed in the barrel muon spectrometer.  The display shows the simulated decay of a $\pi_{\rm{v}}$ from the MC mass point $m_{H}$~=~140~GeV, $m_{\pi_{\rm{v}}}$~=~20~GeV.}
  \label{fig:barvxreco}
\end{figure}

\begin{figure}[ht!]
  \centering
  \begin{subfigure}[b]{0.49\textwidth}
    \includegraphics[width=\textwidth]{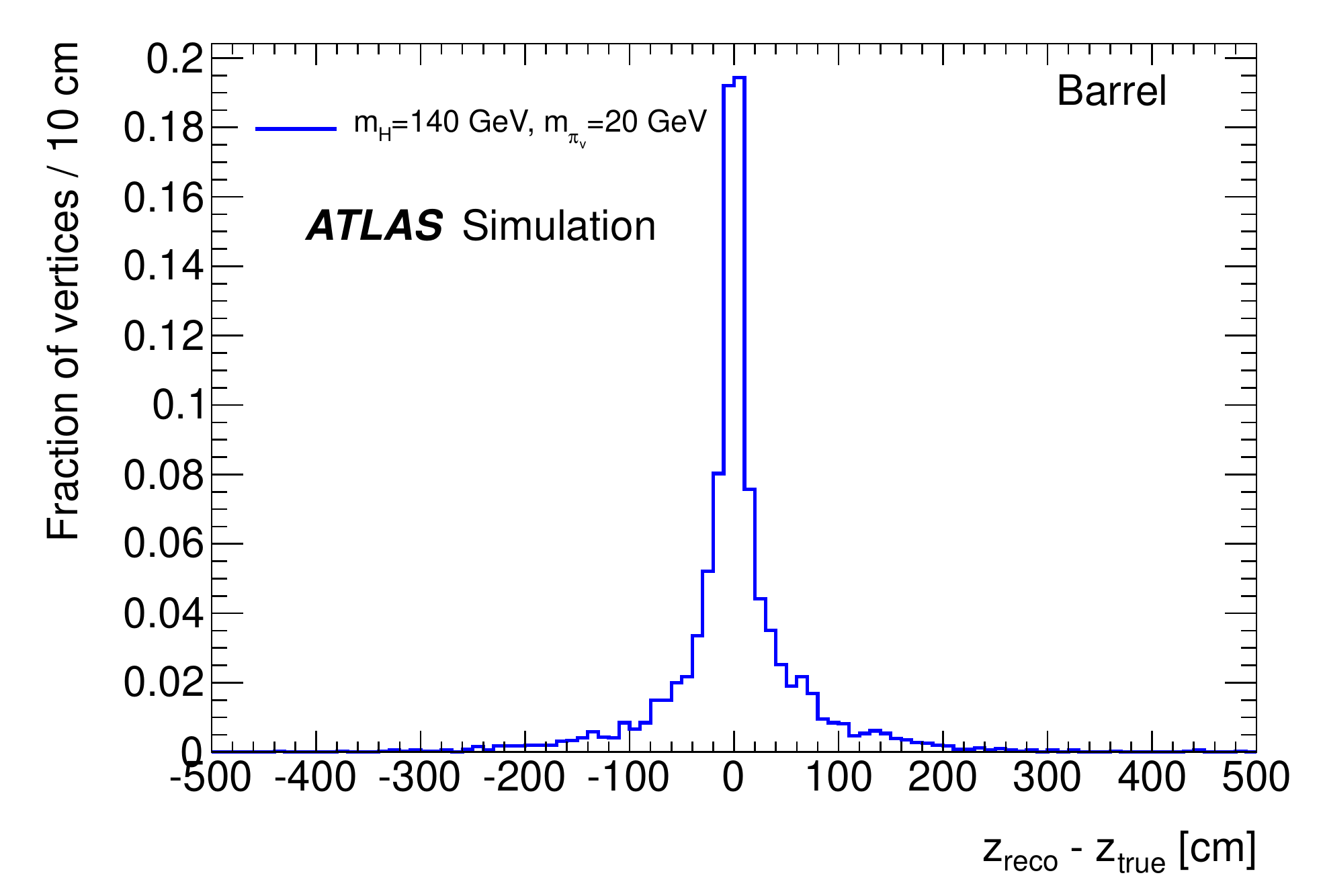}
    \caption{}
  \end{subfigure}
  \begin{subfigure}[b]{0.49\textwidth}
    \includegraphics[width=\textwidth]{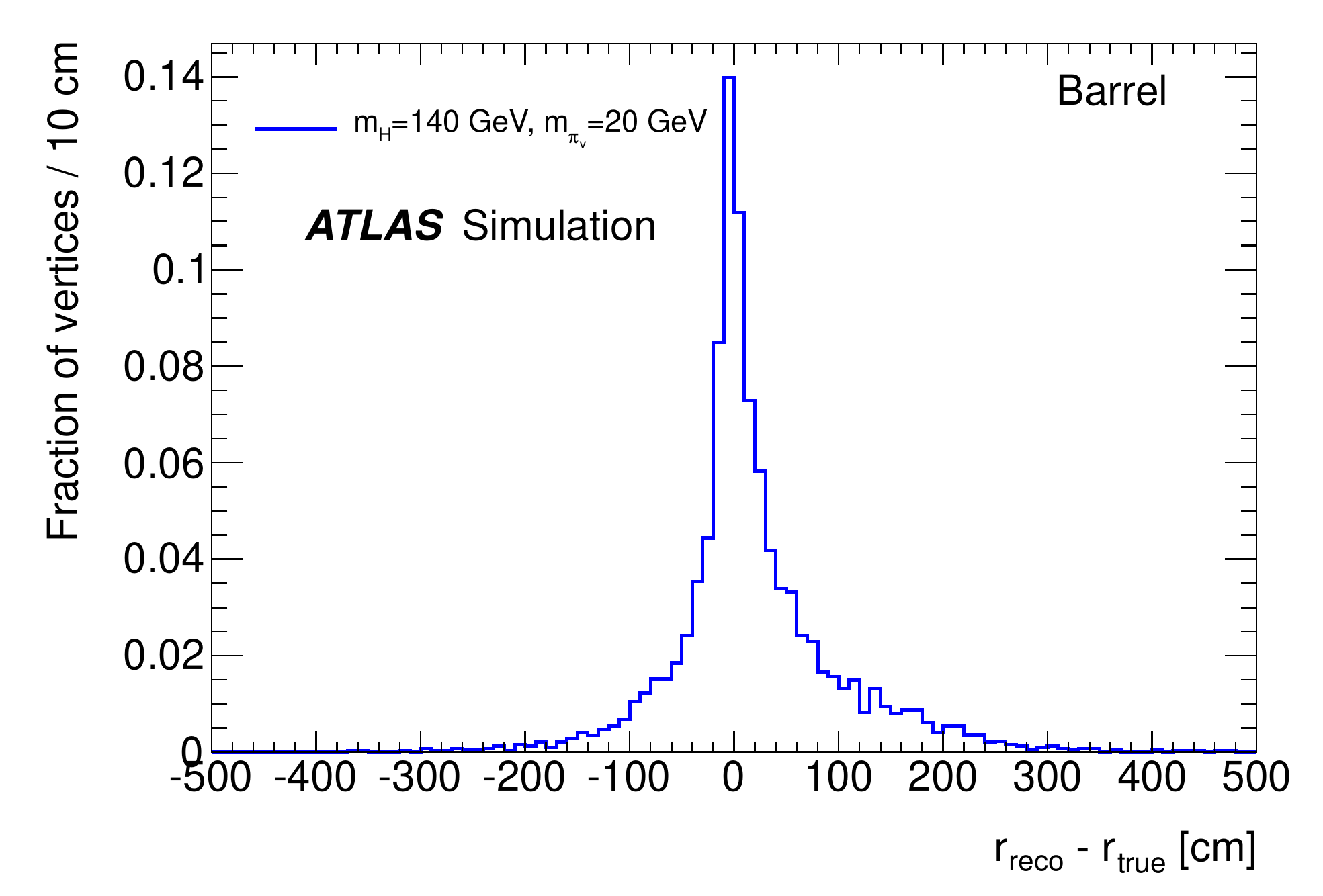}
    \caption{}
  \end{subfigure}
  \caption{Distributions of the vertex-reconstruction residuals in the (a) $z$ coordinate and (b) $r$ coordinate for decays in the barrel MS.  Both plots show the MC mass point $m_{H}$~=~140~GeV, $m_{\pi_{\rm{v}}}$~=~20~GeV.  Similar results are found using the other MC samples.}
  \label{fig:vxresol}
\end{figure}

The barrel vertex-reconstruction algorithm begins by finding the cluster of tracklets to be used in the vertex routine. This is done by using a simple cone algorithm that has a cone of radius $\Delta R$~=~0.6 and its apex at the IP.\footnote{The cone algorithm uses the position of each tracklet as a seed, and the cone containing the most tracklets is chosen as the optimal centre.}  Then the line-of-flight of the decaying particle in the $\theta$ direction is reconstructed by drawing a line between the IP and the centroid of all tracklets in the cluster. Figure~\ref{fig:lineOfFlight}(a) shows the difference between the reconstructed and true $\pi_{\rm{v}}$ line-of-flight in $\theta$.  This method is able to reconstruct the $\theta$ line-of-flight with an RMS of 21~mrad.  The line-of-flight in the $\phi$ direction is computed in two steps. First an approximate $\phi$ line-of-flight is computed by using the $\phi$ coordinate of each tracklet in the cluster to calculate an average $\phi$.\footnote{For each tracklet the $\phi$ coordinate is approximated using the centre of the associated MDT chamber, see section~4.1.}  Then, using this $\phi$ value and the $\theta$ of the $\pi_{\rm{v}}$ line-of-flight, a cone of radius $\Delta R$~=~0.6 with its apex at the IP is constructed and the average of all RPC-phi measurements within this cone is used to determine the $\phi$ line-of-flight.  Figure~\ref{fig:lineOfFlight}(b) shows the difference between the reconstructed and true line-of-flight in $\phi$.  This method is able to reconstruct the $\phi$ line-of-flight with a RMS of 50~mrad, which corresponds to $\sim$1/8 of a large MDT chamber.\\
Due to the lack of precise $\phi$ information for the tracklets and to the inhomogeneous  magnetic field~\cite{ATLASDET} in the MS, it is necessary to perform the vertex reconstruction in a single $r$--$z$ plane. Therefore, the clustered tracklets are all mapped onto the $r$--$z$ plane in which the reconstructed line-of-flight lies.
The tracklets are then back-extrapolated, using the full magnetic field map, in this $r$--$z$ plane to a series of lines parallel to the $z$-axis that are equally spaced along the line-of-flight. The distance, along the line-of-flight, between two adjacent lines is 25~cm and the lines cover the range from $r$~=~3.5~m to $r$~=~7.0~m. This results in 15 lines of constant radius for a line-of-flight at $\eta$~=~0 and 22 lines at $|\eta|$~=~1. Increasing the number of lines in this manner ensures that the vertex routine treats the entire $|\eta|$ range uniformly, by extrapolating each tracklet a constant distance along the line-of-flight.  Figure~\ref{fig:barvxreco} illustrates how these lines of constant radius in the spectrometer are used to back-extrapolate the tracklets and to reconstruct the vertex.\\
The uncertainty arising from the lack of precise $\phi$ information for each tracklet and hence lack of precise knowledge of the magnetic field, is evaluated by rotating the $r$--$z$ plane by 200~mrad around the $z$-axis\footnote{Each large MDT chamber is $\sim$400~mrad wide, thus the rotation of 200~mrad corresponds to the maximum variation of position from the centre of the chamber.}  and again extrapolating the tracklets to the lines of constant radius.  The difference in the $z$ position of the rotated and nominal tracklet is calculated at each line, and taken to be the uncertainty associated with the imprecise knowledge of the magnetic field. This uncertainty is  added, in quadrature, to the position uncertainty arising from the uncertainty in the measured momentum of the tracklet. For tracklets that do not have a momentum measurement, the only source of uncertainty arises from the uncertainty in the tracklet direction, $\alpha$.  Only those tracklets with a total uncertainty $\sigma_{z}$~$<$~20~cm are used by the vertex-finding algorithm.\\
At each line of constant radius, the average $z$ position of the tracklets is computed by weighting the extrapolated tracklet position by 1/$\sigma_{z}^{2}$.  The $\chi^2$ of this candidate vertex (whose $z$ coordinate is assumed to be the average $z$ position) is computed, assuming that the tracklets originate from the vertex point.  If the $\chi^2$ probability for the vertex point is less than 5$\%$, the tracklet with the largest contribution to the total vertex $\chi^2$ is dropped and the vertex point is recomputed.  This is done iteratively, until there is either an acceptable vertex, with $\chi^2$ probability greater than 5$\%$, or there are fewer than three tracklets left to compute the vertex point.\\
The reconstructed vertex position is taken to be the radius and the $z$ position on the line that had the largest number of tracklets used to create the candidate vertex point.  If there are two (or more) lines of constant radius with the same number of tracklets, the line with the minimum $\chi^2$ is selected as the reconstructed vertex position.  The difference in position between the reconstructed and true decay vertex are shown in figure~\ref{fig:vxresol}.
\subsection{Vertex reconstruction in the endcap MS}
\label{sec:endvxreco}
\begin{figure}[ht!]
  \centering
  \begin{subfigure}[b]{0.49\textwidth}
    \includegraphics[width=\textwidth]{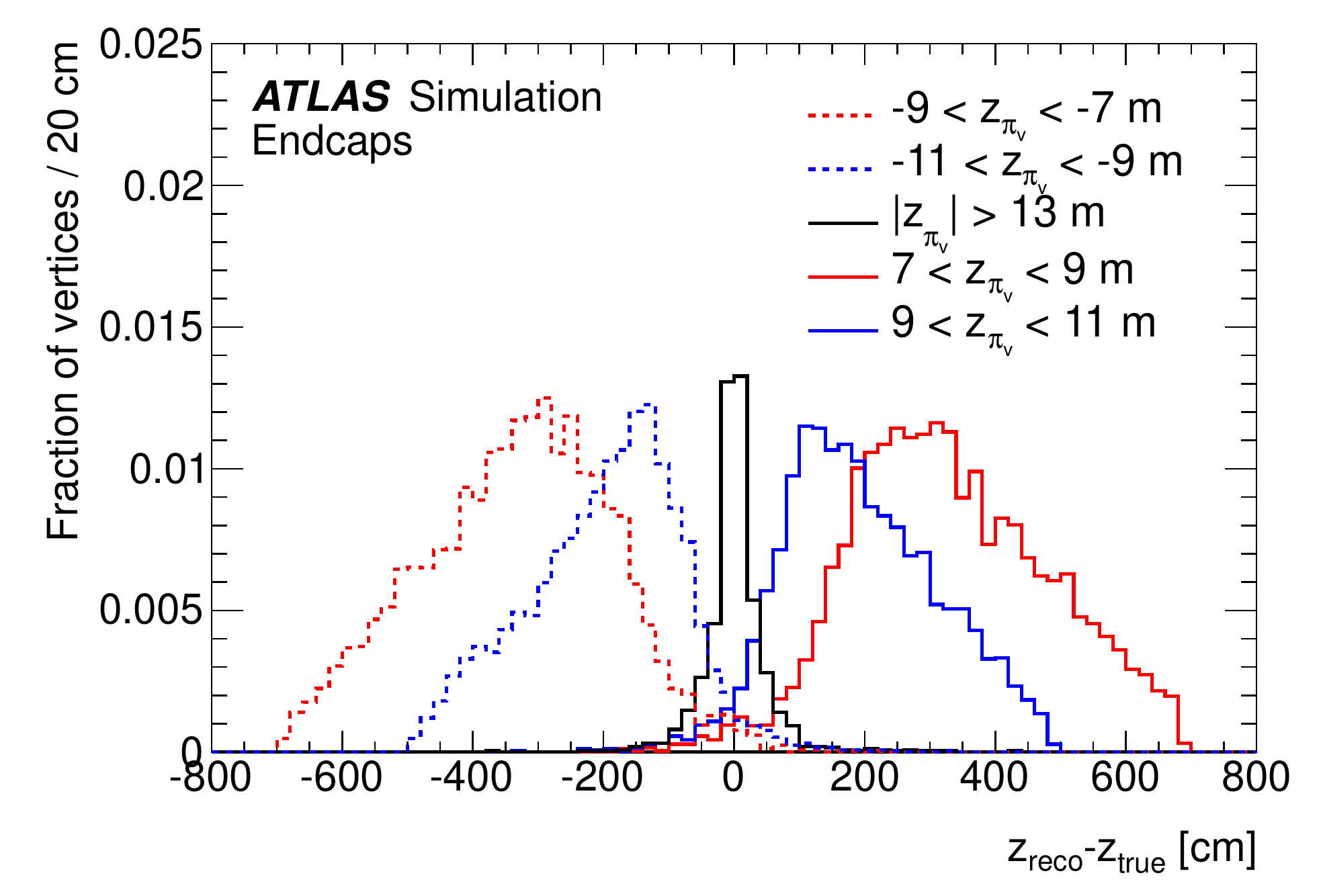}
    \caption{}
  \end{subfigure}
  \begin{subfigure}[b]{0.49\textwidth}
    \includegraphics[width=\textwidth]{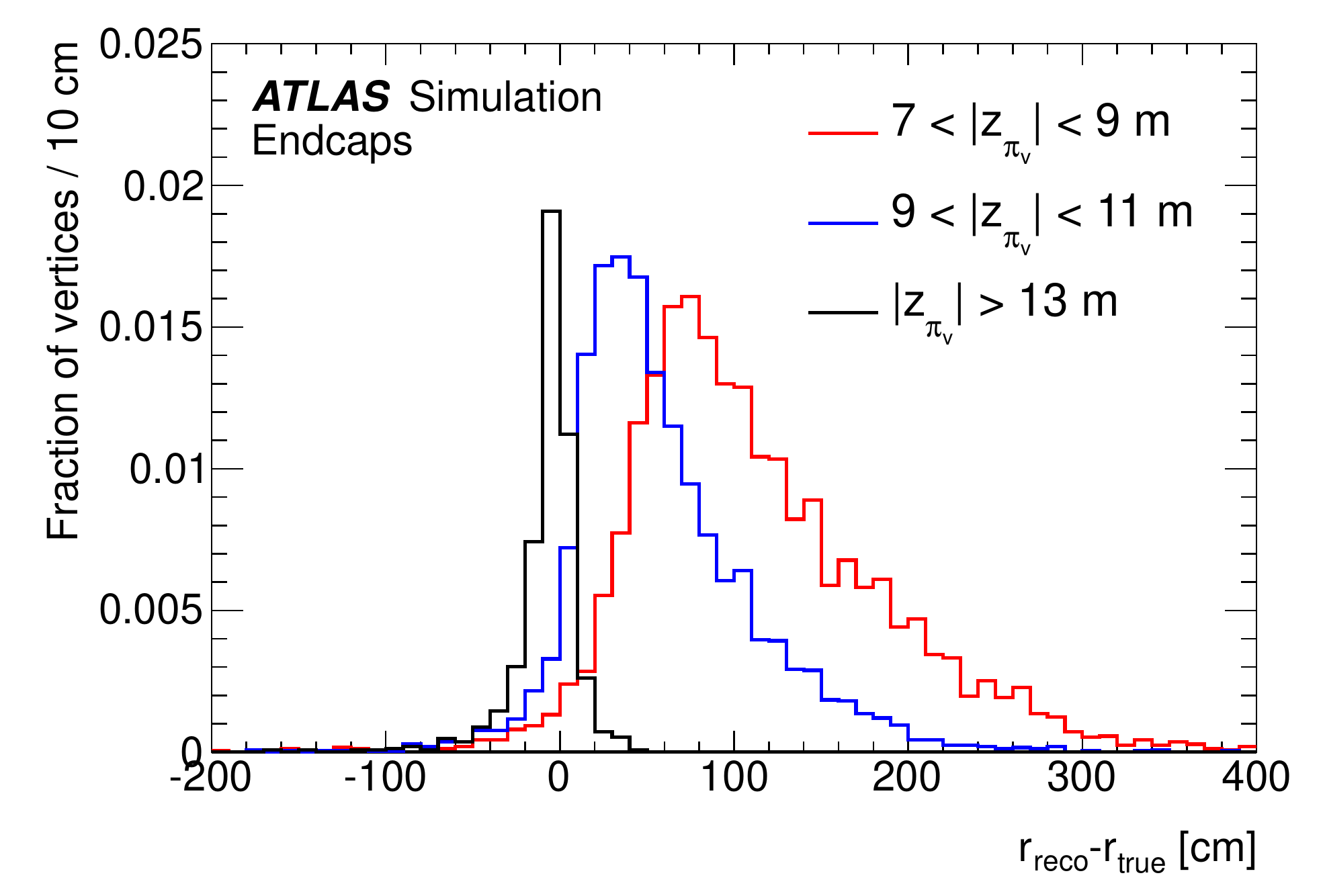}
    \caption{}
  \end{subfigure}
  \caption{Distributions of the vertex-reconstruction residuals in the (a) $z$ coordinate and (b) $r$ coordinate for decays in the endcap MS for different ranges of $z_{\pi_{\rm{v}}}$.  Both plots show the MC mass point $m_{H}$~=~140~GeV, $m_{\pi_{\rm{v}}}$~=~20~GeV.  Similar results are found using the other MC samples.}
  \label{fig:resVpos}
\end{figure}

\begin{figure}[ht!]
  \centering
  \includegraphics[width=125mm]{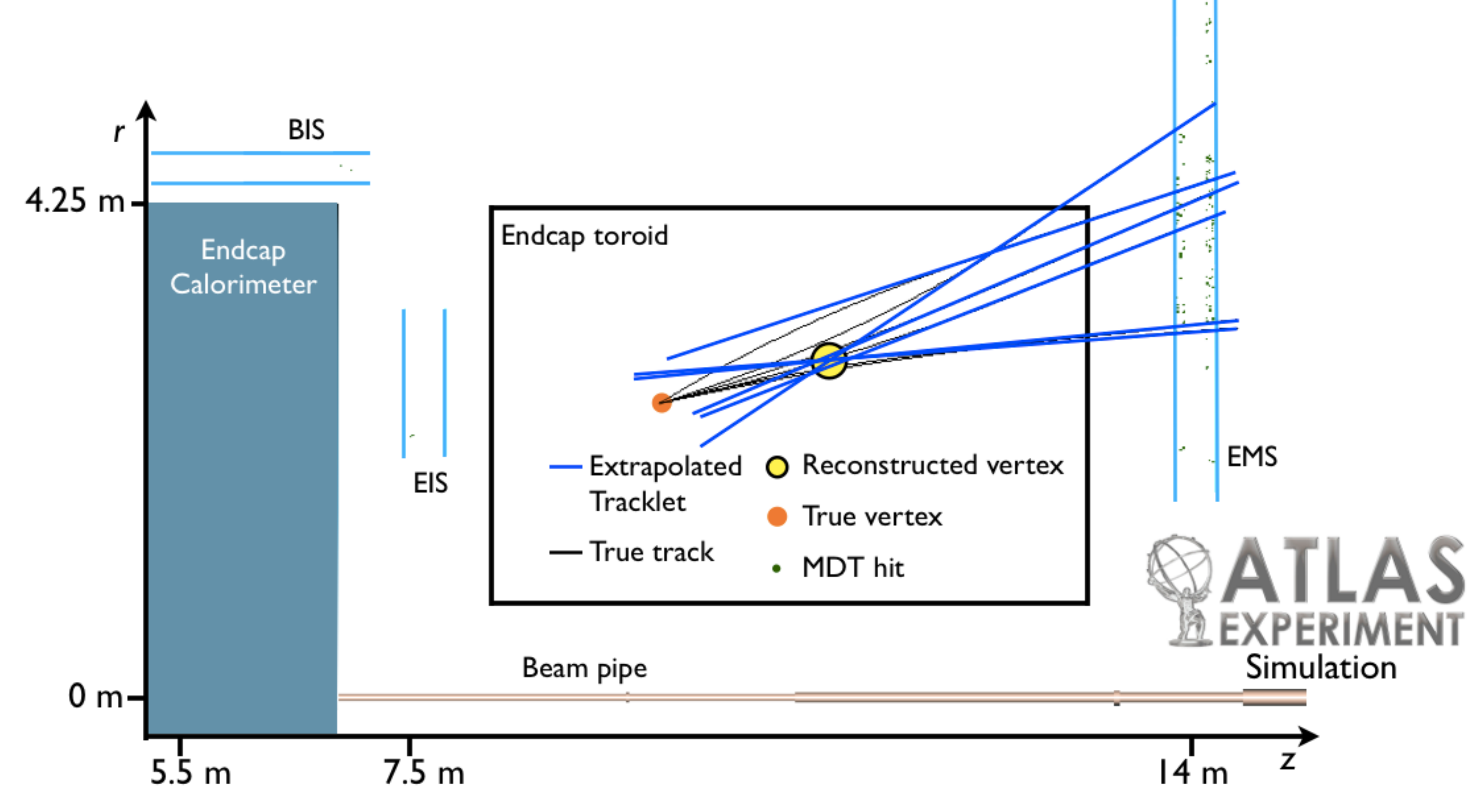}
  \caption{Diagram illustrating the reconstruction technique employed in the endcap muon spectrometer.  The display shows the simulated decay of a $\pi_{\rm{v}}$ from the MC mass point $m_{H}$~=~140~GeV, $m_{\pi_{\rm{v}}}$~=~20~GeV.}
  \label{fig:ecvxreco}
\end{figure}

Tracklets reconstructed in the endcap region (1~$<$~$|\eta|$~$<$~2.7) have no momentum or charge measurements; thus a different approach to vertex finding is required.  As discussed in section~\ref{sec:msdecays}, detectable decays occur just upstream of or inside the endcap toroid. As a consequence, measurements of the charged particles' trajectories are made after they have been bent by the magnetic field.  This implies that the tracklets will need to be back-extrapolated as straight lines into the endcap toroid.  Therefore, in the endcap MS, a simple linear extrapolation and minimization routine is used to reconstruct the decay vertices.  The routine starts by grouping the tracklets that are clustered in ($\eta$,$\phi$), using the same cone algorithm that is employed in the barrel vertex-reconstruction routine.  The lines-of-flight in $\theta$ and $\phi$ are calculated as in the barrel, except the TGC-phi measurements are used instead of the RPC-phi measurements.  The resolution in the lines-of-flight in both $\theta$ and $\phi$ is comparable to the resolution achieved in the barrel using the simulated signal samples.\\  
The clustered tracklets provide constraining equations of the form \mbox{$\beta$$_i$~=~$-r$~tan$\alpha_i$ + $z$}, where $\beta_{i}$ is the $z$-intercept and $\alpha_{i}$ the angle of the $i$-th tracklet, which are used in a least squares regression fit of the vertex.  The vertex position is then iterated, dropping the tracklet that is farthest from the vertex until the distance of closest approach between the farthest tracklet and the vertex is less than 30~cm.  The vertex position is accepted if it is reconstructed using at least three tracklets, is within the endcap MS fiducial volume, and is upstream of the middle station ($|z|$~=~14~m).  Figure~\ref{fig:resVpos} shows the position of the reconstructed vertices with respect to the true decay point.  The magnetic field in the endcap toroid bends the charged tracks while preserving the line-of-flight of the neutral long-lived particle.  Because the tracklets are measured after the magnetic field and extrapolated back into the magnetic field region as straight lines, the vertex position is systematically shifted to larger values of $|z_{\rm{reco}}|$ with respect to the true decay position.  Due to the line-of-flight being preserved by the magnetic field, the reconstructed vertices are also shifted to larger values of $r_{\rm{reco}}$.  This effect lessens as the decay occurs closer to the outer edge of the endcap toroid ($|z|$~$\sim$~12.5~m) and the charged particles experience less bending making the straight-line extrapolation used in the vertex reconstruction a better approximation. Figure~\ref{fig:ecvxreco} illustrates this reconstruction technique and the systematic shifts in both $r_{\rm{reco}}$ and $|z_{\rm{reco}}|$. 

\section{Performance}
\label{sec:goodvx}
The performance of the vertex-reconstruction algorithms has been evaluated on both data and MC simulation.  Data events corresponding to 1.94~fb$^{-1}$ collected in 2011 at $\sqrt{s}$~=~7~TeV were analysed.  The events in both data and MC simulation were required to pass the Muon RoI Cluster trigger.  Additionally, the data events were required to have been collected during a period when all detector elements were operational.  
\subsection{Good vertex selection}
\label{sec:gvs}
\begin{figure}[htb!]
  \centering
  \begin{subfigure}[b]{0.49\textwidth}
    \includegraphics[width=\textwidth]{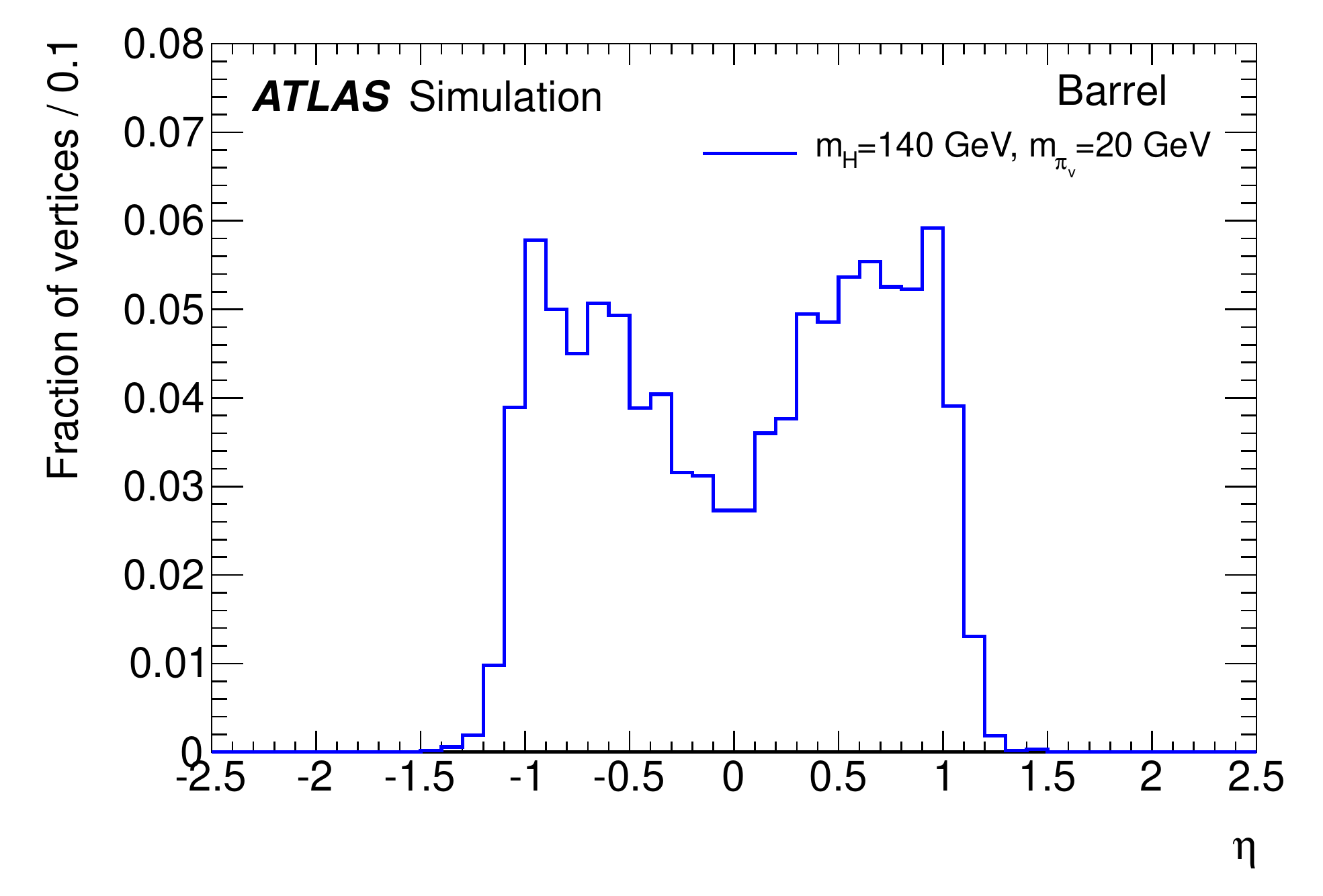}
    \caption{}
  \end{subfigure}
  \begin{subfigure}[b]{0.49\textwidth}
    \includegraphics[width=\textwidth]{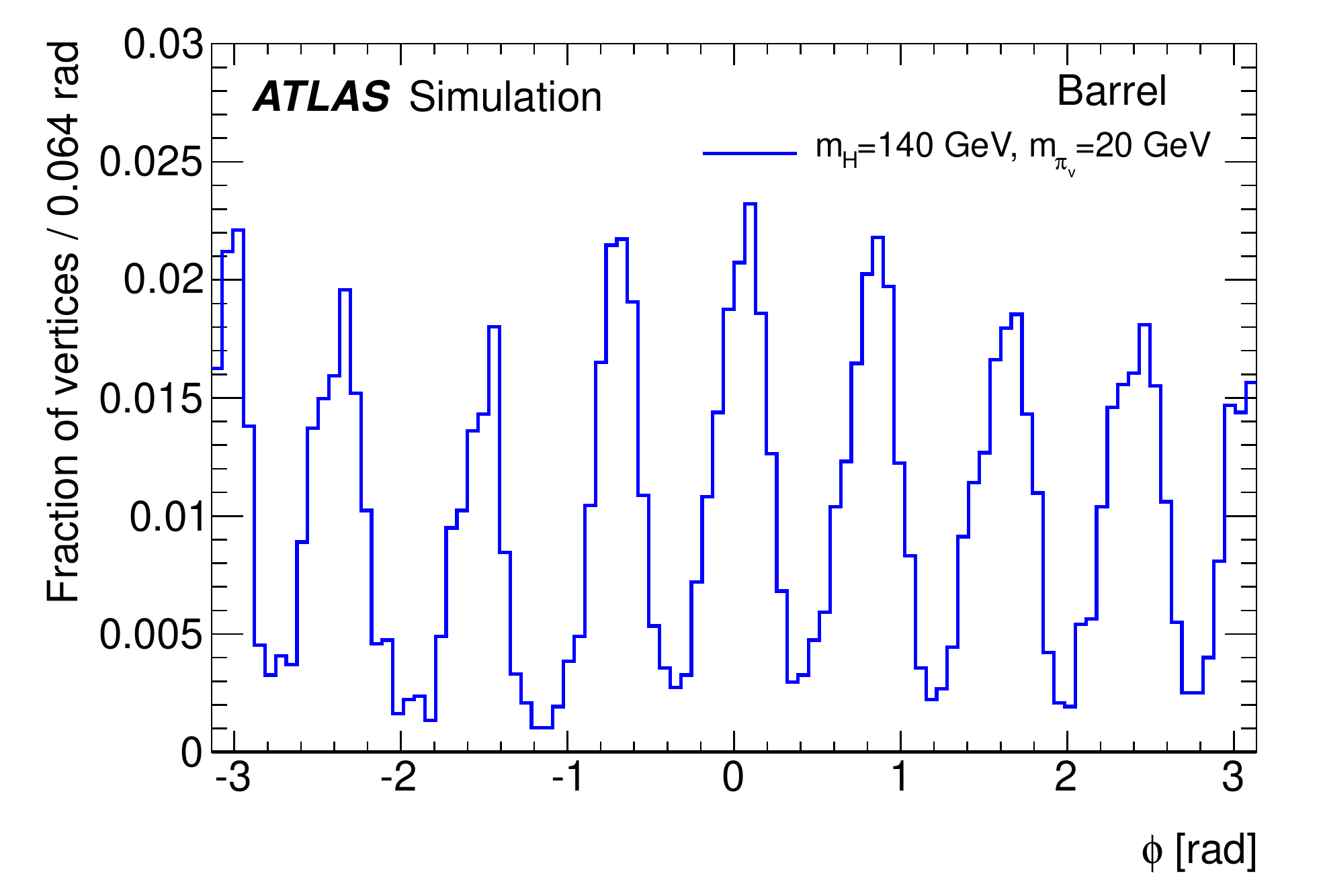}
    \caption{}
  \end{subfigure}
  \caption{Distribution of the (a) $\eta$ and (b) $\phi$ positions of the good vertices in the barrel muon spectrometer for the Monte Carlo signal samples.  Both plots show the MC mass point $m_{H}$~=~140~GeV, $m_{\pi_{\rm{v}}}$~=~20~GeV.  Similar results are found using the other MC samples.}
  \label{fig:mcvxpos}
\end{figure}

Events with vertices that originate from detector noise, cosmic ray showers or punch-through hadronic jets\footnote{A punch-through hadronic jet  occurs when particles from a hadronic jet escape the calorimeter volume.} can be rejected by imposing a series of selection criteria.  Vertices found in the barrel MS are required to be consistent with the decay of a long-lived particle that originates at the IP. Therefore the sum of the $p_{z}$ of all tracklets used in the vertex fit is required to point away from the IP.\footnote{The pointing requirement is only applied to vertices in the region $|\eta|$~$>$~0.3 to avoid inefficiencies due to the momentum resolution of the tracklets in the region where the sum of $p_z$ is expected to be $\sim$0.}  Because the MS tracklets in the endcaps do not have an associated momentum measurement, it is not possible to extrapolate them through the endcap toroids.  Therefore, the pointing requirement is only applied to vertices reconstructed in the barrel MS.  The vertex is required to be in a region with high activity in the MDT and trigger chambers.  To remove events with coherent noise in the MDTs, the vertex is required to be in a region of the detector with fewer than 3000 MDT hits and be reconstructed with tracklets from at least two different muon stations.  Additionally, vertices can be required to be isolated from ID tracks and/or hadronic jets to reduce backgrounds and punch-through contamination.  These isolation criteria are analysis dependent and therefore beyond the scope of this paper.\\
In order for a vertex to be considered good, the following criteria must be satisfied:
\begin{itemize}
  \item Tracklets: The vertex must contain tracklets reconstructed from at least two different muon stations (i.e. inner + middle, middle + outer, inner + outer or small sector + large sector).
  \item Vertex Pointing (barrel MS only): The sum of the momenta of all tracklets used in the vertex fit is required to point away from the IP ($\sum\limits_{\rm{tracklets}}$$p_{z}^{\rm{tracklet}}$$\cdot$$z_{\rm{vertex}}$~$>$~0).
  \item $N_{\rm{MDT}}$: The number of MDT hits contained in a cone of $\Delta R$~$=$~0.6 around the vertex is required to be in the range 200~$<$~$N_{\rm{MDT}}$~$<$~3000.
  \item $N_{\rm{RPC (TGC)}}$: The number of RPC-eta (TGC-eta) hits contained in a cone of $\Delta R$~$=$~0.6 is required to be $N_{\rm{RPC (TGC)}}$~$>$~100 for vertices reconstructed in the barrel (endcap) MS.
\end{itemize}
Figure~\ref{fig:mcvxpos} shows distributions of the position of the good vertices in the ATLAS coordinate system for MC simulation. The depletion of vertices near $\eta$~=~0 is due to the limited coverage of the RPC trigger chambers in this region and hence limited coverage of the Muon RoI Cluster trigger.  Figure~\ref{fig:mcvxpos}(b) shows that in the barrel MS the algorithm has a much lower efficiency near the magnet coils ($\phi\sim\pm(2n+1)\mathrm{\pi}/4$).  Decays occurring in the region near the coils produce a larger number of noise hits, due to the large amount of material present in the magnet coil, which lowers the tracklet reconstruction efficiency (see figure~\ref{fig:ntrackletphi}). The effects of the reduced MDT coverage and the extra material present in the region of the feet supporting the detector is visible as a relative decrease in the number of reconstructed vertices in the region between $\phi$~=~$-$1 and $\phi$~=~$-$2.\\
 The tracklets reconstructed in the BIS and BOS chambers, which are located near the magnet coils, do not have a momentum measurement, and these tracklets are extrapolated through the magnetic field as straight lines. Therefore BIS and BOS tracklets may have a large uncertainty in their extrapolation and are often rejected by the $\chi^2$ test described in section~\ref{sec:barvxreco}.
\subsection{Vertex reconstruction efficiency}
\begin{figure}[htb]
  \centering
  \begin{subfigure}[b]{0.49\textwidth}
    \includegraphics[width=\textwidth]{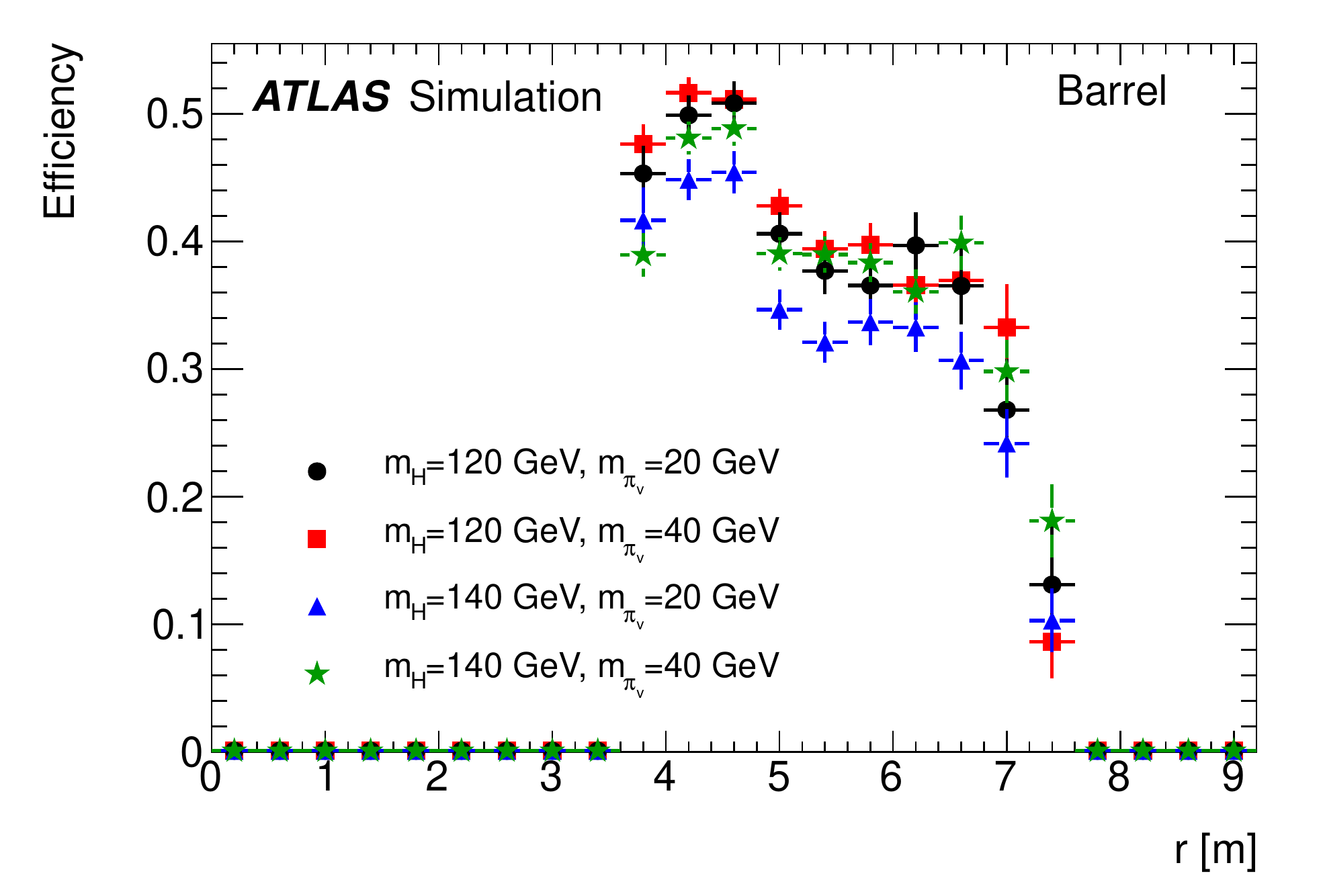}
    \caption{}
  \end{subfigure}
  \begin{subfigure}[b]{0.49\textwidth}
    \includegraphics[width=\textwidth]{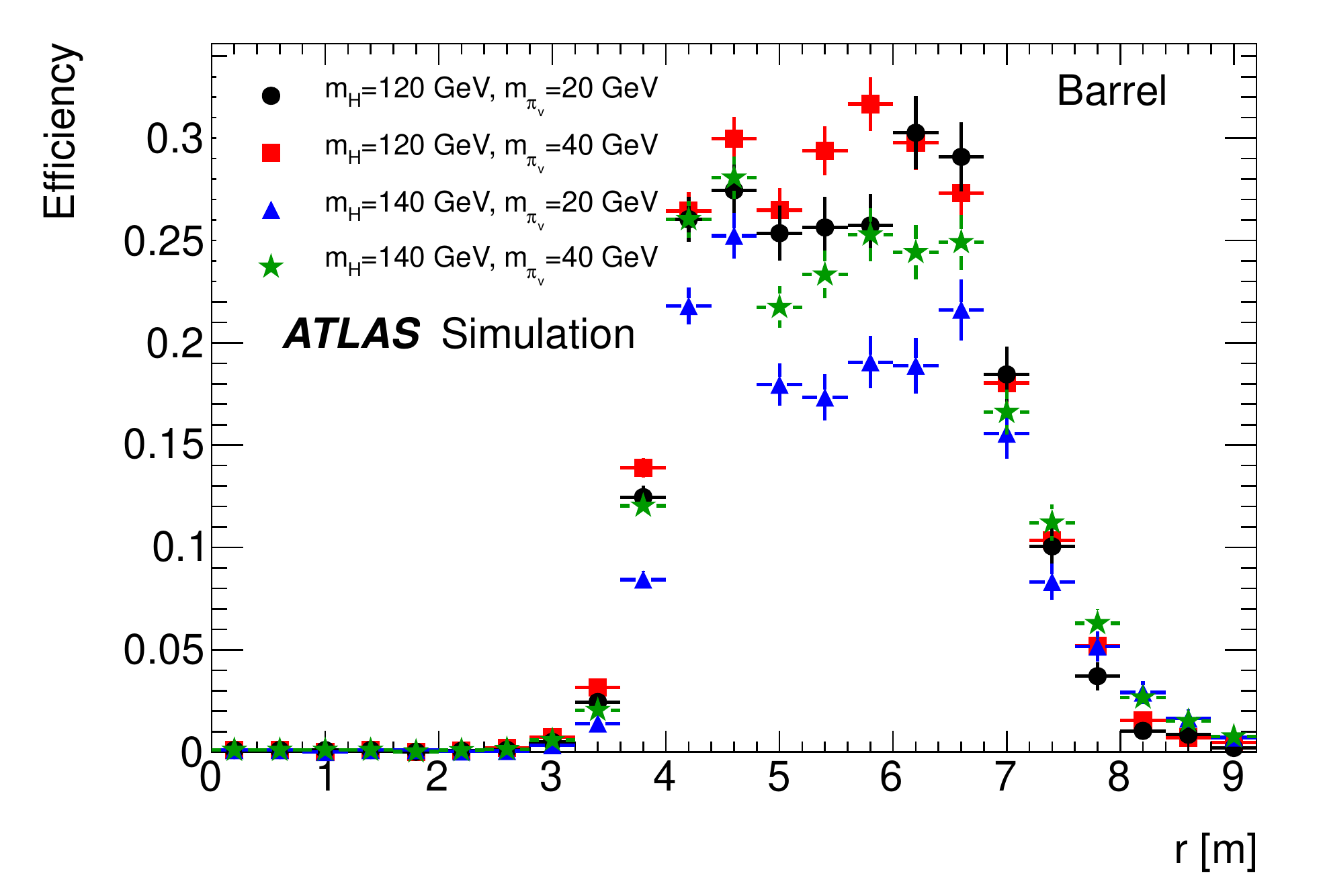}
    \caption{}
  \end{subfigure}
  \caption{Efficiency for reconstructing a vertex for $\pi_{\rm{v}}$ decays in the barrel muon spectrometer as a function of the radial decay position of the $\pi_{\rm{v}}$ for (a) $\pi_{\rm{v}}$ decays that satisfy the Muon RoI Cluster trigger and (b) $\pi_{\rm{v}}$ decays that do not satisfy the Muon RoI Cluster trigger.}
  \label{fig:msvxBarEff}
\end{figure}

\begin{figure}[htb]
  \centering
  \includegraphics[width=100mm]{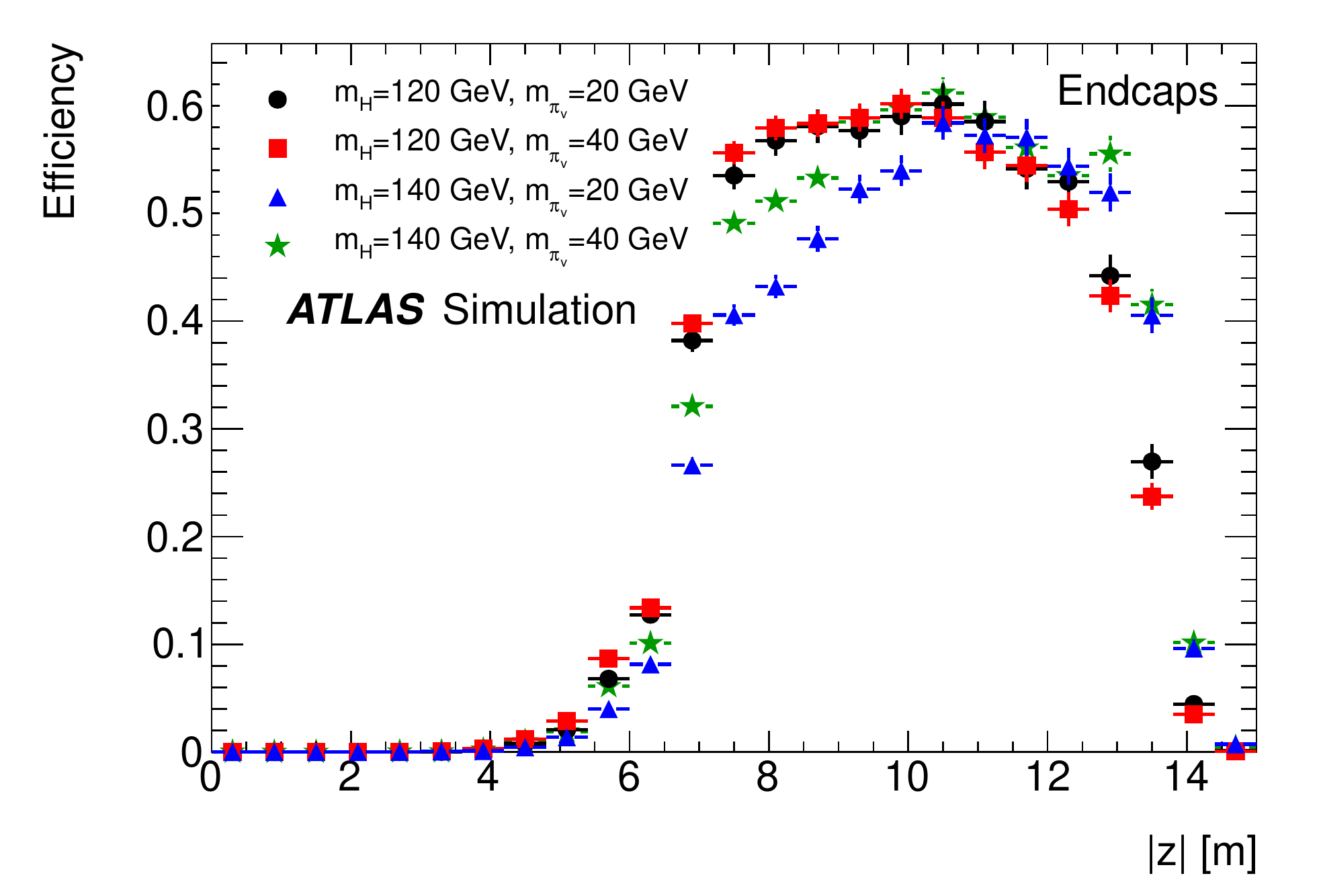}
  \caption{Efficiency for reconstructing a vertex for $\pi_{\rm{v}}$ decaying in the endcap MS as a function of the absolute $z$ position of the $\pi_{\rm{v}}$ decay.}
  \label{fig:msvxECEff}
\end{figure}

The efficiency for vertex reconstruction is defined as the fraction of simulated $\pi_{\rm{v}}$ decays occurring in the MS fiducial volume that have a reconstructed vertex satisfying all of the criteria described in section~\ref{sec:gvs}.  Figure~\ref{fig:msvxBarEff}(a) shows the efficiency to reconstruct a vertex in the barrel MS for $\pi_{\rm{v}}$ decays that satisfy the Muon RoI Cluster trigger requirements and figure~\ref{fig:msvxBarEff}(b) the efficiency for those $\pi_{\rm{v}}$ decays that do not satisfy the trigger.  When a $\pi_{\rm{v}}$ satisfies the trigger requirements, the vertex-reconstruction efficiency varies from $\sim$50$\%$ near the calorimeter face ($r$~$\sim$~4~m) to $\sim$30$\%$ for decays occurring close to the middle station ($r$~$\sim$~7~m). The efficiency decreases as the decay occurs closer to the middle station because the charged hadrons and photons (and their corresponding EM showers) have not spatially separated and are overlapping when they traverse the middle station.  This reduces the reconstruction efficiency for tracklet reconstruction and consequently for vertex reconstruction.  Those $\pi_{\rm{v}}$ that do not satisfy the Muon RoI Cluster trigger requirements have a lower reconstruction efficiency because these decays tend to have all of their decay products entering a single sector. Therefore the tracks and EM showers are often overlapping and the tracklet recontruction efficiency is lower, which in turn reduces the vertex reconstruction efficiency.
\\The efficiency for reconstructing vertices in the endcaps as a function of $|z|$, shown in figure~\ref{fig:msvxECEff}, is roughly constant from 7~m to 14~m and varies between 45$\%$ and 60$\%$ depending on the signal model parameters.  
\subsection{Performance on 2011 collision data}
\begin{figure}[htb!]
  \centering
  \begin{subfigure}[b]{0.49\textwidth}
    \includegraphics[width=\textwidth]{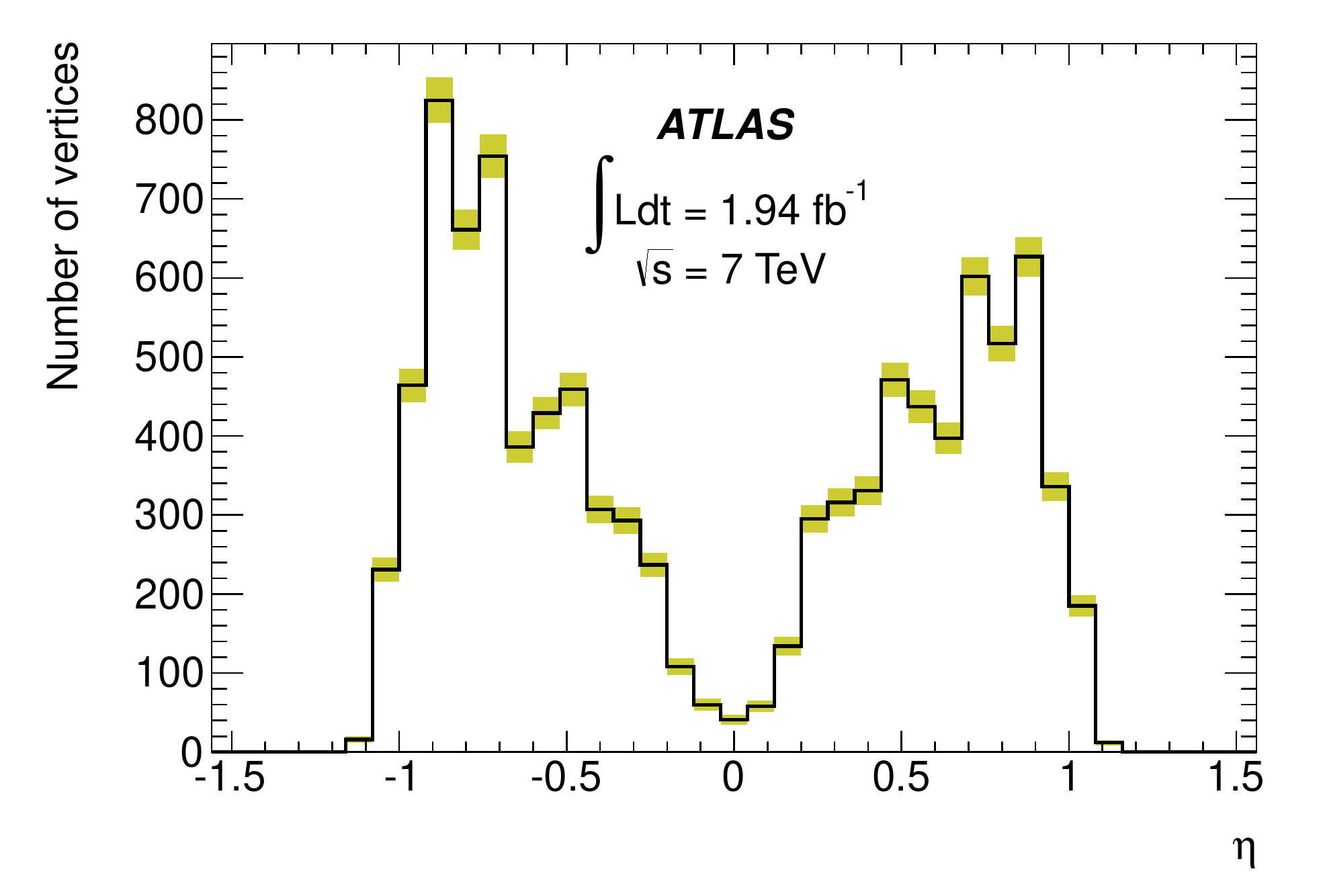}
    \caption{}
  \end{subfigure}
  \begin{subfigure}[b]{0.49\textwidth}
    \includegraphics[width=\textwidth]{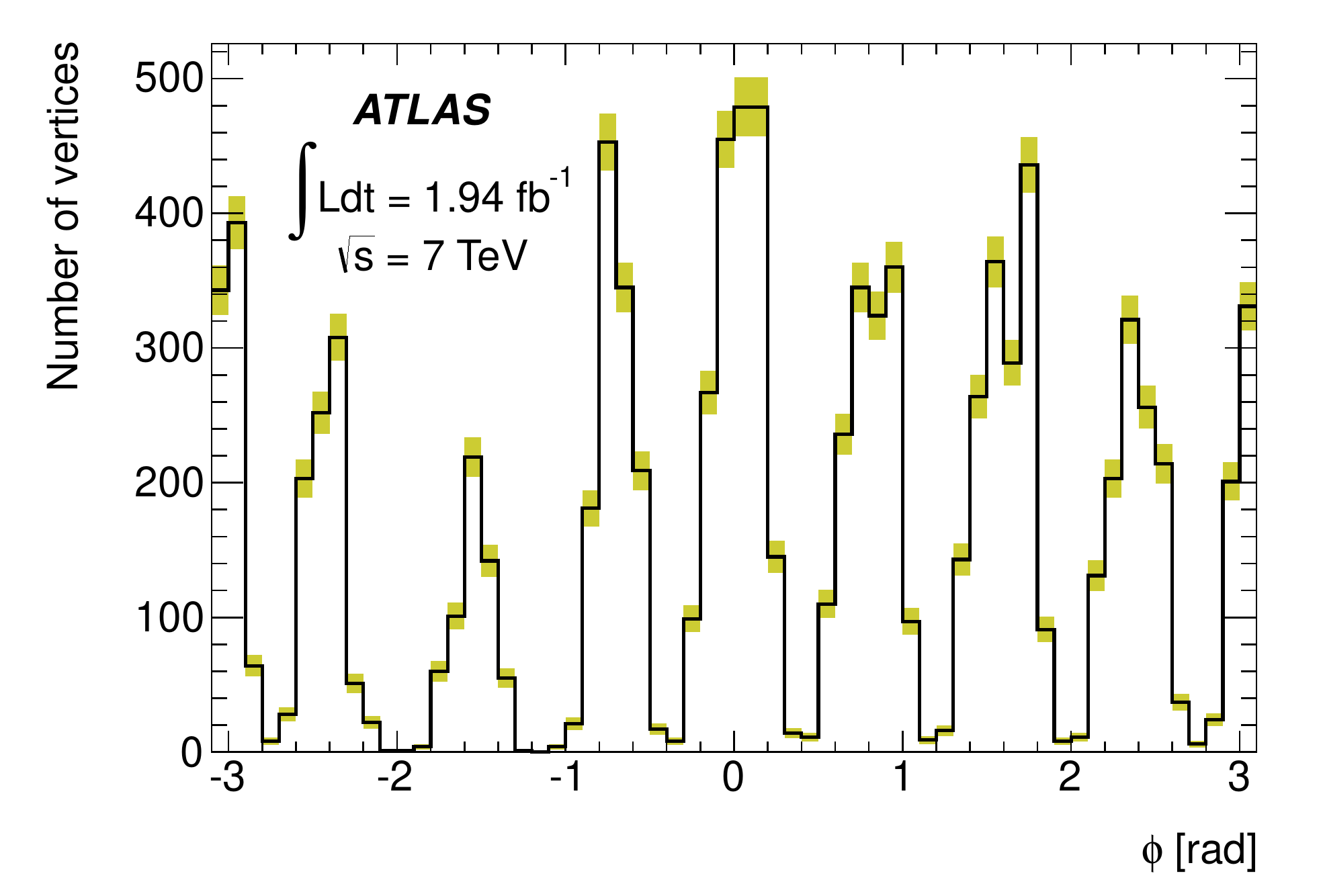}
    \caption{}
  \end{subfigure}
  \caption{Distribution of the (a) $\eta$ and (b) $\phi$ position of the good vertices reconstructed in the data events that pass the Muon RoI Cluster trigger.  The statistical uncertainty on the data points is represented by the yellow bands. The effect of extra material present in the feet is visible as a relative decrease in the number of vertices reconstructed in the region between $\phi$~=~$-$1 and $\phi$~=~$-$2. }
  \label{fig:datavxpos}
\end{figure}

\begin{figure}[htb!]
  \centering
  \includegraphics[width=100mm]{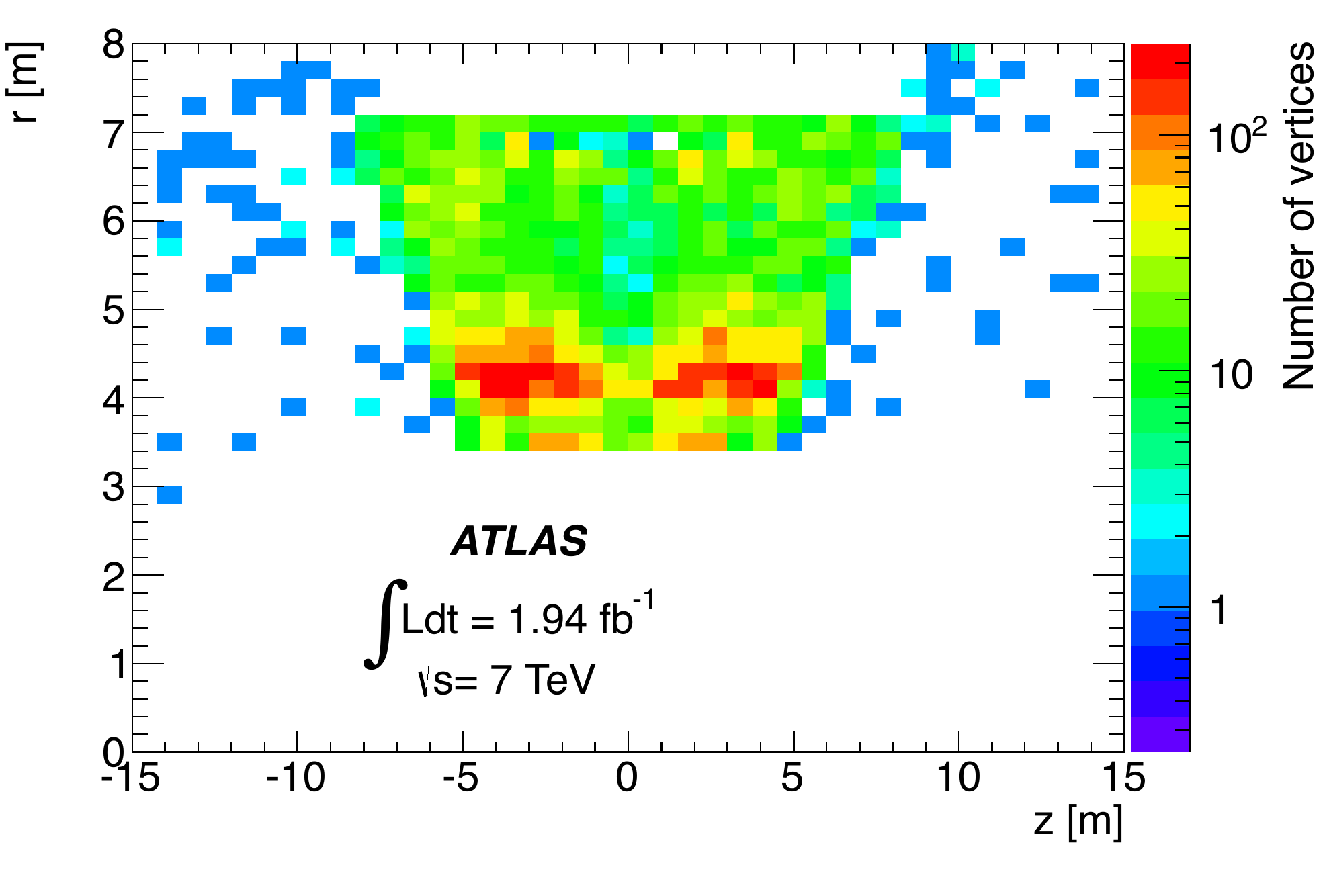}
  \caption{Distribution of the position, in ($r$,$z$), of all good vertices in the fiducial volume of the muon spectrometer in events that passed the barrel Muon RoI Cluster trigger from 1.94~fb$^{-1}$ of 7~TeV $pp$ collision data. }
  \label{fig:datavxRZ}
\end{figure}

In 2011, the Muon RoI Cluster trigger was only active in the barrel MS region; therefore, no events with a vertex in the endcaps are found using this selection.  The position of the vertices found in $\eta$ and $\phi$, after applying the criteria for a good vertex, are shown in figure~\ref{fig:datavxpos} for those vertices reconstructed in 1.94~fb$^{-1}$ of 2011 data.  The vertices found in data show similar features in $\eta$ and $\phi$ as in the simulation (figure~\ref{fig:mcvxpos}).\\
The position in ($r$,$z$) of the good vertices reconstructed in data is shown in figure~\ref{fig:datavxRZ}.  The vertices tend to group around $|z|$~$\sim$~4~m and $r$~$\sim$~4~m, near the gap region between the barrel and extended barrel calorimeters where particles can escape the calorimeter volume. The few events with a vertex in the endcap MS are due to events with activity in both the barrel and endcap MS. The activity in the barrel MS satisfies the Muon RoI Cluster trigger, while the activity in the endcaps produces the reconstructed vertex.
\subsection{Data--Monte Carlo simulation comparison}
\begin{figure}[ht!]
  \centering
  \includegraphics[width=100mm]{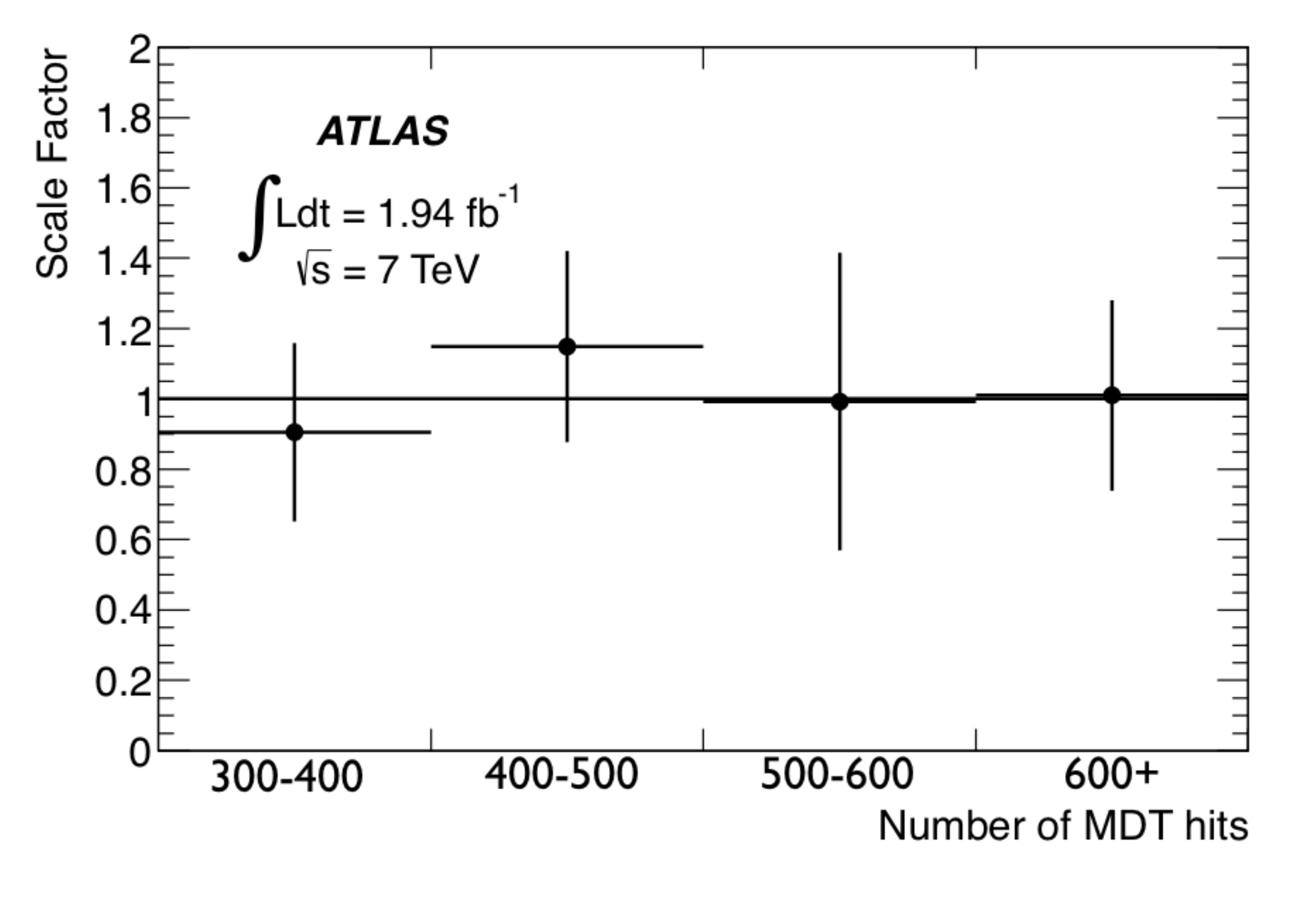}
  \caption{The scale factor calculated by dividing the fraction of punch-through jets in data that have a good MS barrel vertex by the fraction in Monte Carlo simulation.  The scale factor is calculated as a function of the number of MDT hits inside a cone of size $\Delta R$~=~0.6, centred around the jet axis.}
  \label{fig:vxcompare}
\end{figure}

To verify the performance of the MC simulation, collision data need to be compared with the MC simulation. This data--MC simulation comparison is analysis dependent and needs to be done in any analysis that makes use of this reconstruction algorithm.  The systematic uncertainty associated with the vertex reconstruction algorithm will, in general, depend upon the event selection and vertex criteria employed in the analysis.  A 2011 analysis searching for pair production of particles decaying in the MS~\cite{HVPRL} performed a study comparing the response of the MS to punch-through jets in data and MC simulation.  The punch-through events are similar to the signal events as they both contain low-energy photons and charged hadrons in a narrow region of the MS.  The candidate punch-through jets were selected from a sample of events that passed a single-jet trigger. In addition, the punch-through jet was required to be in the barrel region of the calorimeter \mbox{($|\eta|$~$<$~1.5)}, satisfy \mbox{$p_{\rm{T}}^{\rm{jet}}$~$>$~20~GeV} and contain a minimum of 300 MDT hits in a cone of $\Delta R$~=~0.6, centred around the jet axis.  To verify that the jet was produced in the collision and not related to machine noise or cosmic rays, the jet was required to contain at least four tracks with $p_{\rm{T}}$~$>$~1~GeV in the ID. The analysis then compared the fraction of punch-through jets that produced a vertex in the barrel MS as a function of the number of MDT hits in a cone of $\Delta R$~=~0.6 centred around the jet axis.  Figure~\ref{fig:vxcompare} shows the ratio of data to Monte Carlo simulation for the fraction of punch-through jets that have a good vertex.  The ratio, within uncertainty, is constant for all numbers of MDT hits in the jet cone. A straight line fit to these data points yields a constant of 1.01~$\pm$~0.15. Therefore, the MC modelling is shown to reproduce data to within 15$\%$ and is independent of the number of MDT hits in the jet cone.  This comparison was performed only in the barrel region, where there was trigger coverage during the 2011 data-taking period.  

\section{Conclusions}
In this paper a new algorithm to reconstruct multi-particle vertices inside the ATLAS muon spectrometer has been presented. This algorithm is able to reconstruct vertices with good efficiency in high-occupancy environments due to electromagnetic showers from $\pi^0$ decays. It can be used in searches for long-lived particles that decay to several charged and neutral particles in the muon spectrometer.  The algorithm has been evaluated on a sample of simulated Higgs boson events in which the Higgs boson decays to long-lived neutral particles that in turn decay to $b\overline{b}$ final states.  Using this benchmark model, the algorithm is found to have an efficiency of $\sim$30--50$\%$ in the barrel muon spectrometer and $\sim$45--60$\%$ in the endcap muon spectrometer.  The performance of the algorithm is also evaluated on 1.94~fb$^{-1}$ of $pp$ collision data at $\sqrt{s}$~=~7~TeV collected in ATLAS during the 2011 data-taking period at the LHC.  A comparison between punch-through jets in data and Monte Carlo simulation shows that the algorithm is performing with good efficiency even in cases of high-chamber occupancy due to electromagnetic showers from $\pi^0$ decays.


\section*{Acknowledgements}

We thank CERN for the very successful operation of the LHC, as well as the
support staff from our institutions without whom ATLAS could not be
operated efficiently.

We acknowledge the support of ANPCyT, Argentina; YerPhI, Armenia; ARC,
Australia; BMWF and FWF, Austria; ANAS, Azerbaijan; SSTC, Belarus; CNPq and FAPESP,
Brazil; NSERC, NRC and CFI, Canada; CERN; CONICYT, Chile; CAS, MOST and NSFC,
China; COLCIENCIAS, Colombia; MSMT CR, MPO CR and VSC CR, Czech Republic;
DNRF, DNSRC and Lundbeck Foundation, Denmark; EPLANET, ERC and NSRF, European Union;
IN2P3-CNRS, CEA-DSM/IRFU, France; GNSF, Georgia; BMBF, DFG, HGF, MPG and AvH
Foundation, Germany; GSRT and NSRF, Greece; ISF, MINERVA, GIF, DIP and Benoziyo Center,
Israel; INFN, Italy; MEXT and JSPS, Japan; CNRST, Morocco; FOM and NWO,
Netherlands; BRF and RCN, Norway; MNiSW and NCN, Poland; GRICES and FCT, Portugal; MNE/IFA, Romania; MES of Russia and ROSATOM, Russian Federation; JINR; MSTD,
Serbia; MSSR, Slovakia; ARRS and MIZ\v{S}, Slovenia; DST/NRF, South Africa;
MINECO, Spain; SRC and Wallenberg Foundation, Sweden; SER, SNSF and Cantons of
Bern and Geneva, Switzerland; NSC, Taiwan; TAEK, Turkey; STFC, the Royal
Society and Leverhulme Trust, United Kingdom; DOE and NSF, United States of
America.

The crucial computing support from all WLCG partners is acknowledged
gratefully, in particular from CERN and the ATLAS Tier-1 facilities at
TRIUMF (Canada), NDGF (Denmark, Norway, Sweden), CC-IN2P3 (France),
KIT/GridKA (Germany), INFN-CNAF (Italy), NL-T1 (Netherlands), PIC (Spain),
ASGC (Taiwan), RAL (UK) and BNL (USA) and in the Tier-2 facilities
worldwide.

\begin{flushleft}
{\Large The ATLAS Collaboration}

\bigskip

G.~Aad$^{\rm 48}$,
T.~Abajyan$^{\rm 21}$,
B.~Abbott$^{\rm 112}$,
J.~Abdallah$^{\rm 12}$,
S.~Abdel~Khalek$^{\rm 116}$,
O.~Abdinov$^{\rm 11}$,
R.~Aben$^{\rm 106}$,
B.~Abi$^{\rm 113}$,
M.~Abolins$^{\rm 89}$,
O.S.~AbouZeid$^{\rm 159}$
% ...
\bigskip
\\
% ...
$^{11}$ Institute of Physics, Azerbaijan Academy of Sciences, Baku, Azerbaijan\\
$^{12}$ Institut de F{\'\i}sica d'Altes Energies and Departament de F{\'\i}sica de la Universitat Aut{\`o}noma de Barcelona, Barcelona, Spain\\
% ...
$^{21}$ Physikalisches Institut, University of Bonn, Bonn, Germany\\
% ...
$^{48}$ Fakult{\"a}t f{\"u}r Mathematik und Physik, Albert-Ludwigs-Universit{\"a}t, Freiburg, Germany\\
% ...
$^{89}$ Department of Physics and Astronomy, Michigan State University, East Lansing MI, United States of America\\
% ...
$^{106}$ Nikhef National Institute for Subatomic Physics and University of Amsterdam, Amsterdam, Netherlands\\
% ...
$^{112}$ Homer L. Dodge Department of Physics and Astronomy, University of Oklahoma, Norman OK, United States of America\\
$^{113}$ Department of Physics, Oklahoma State University, Stillwater OK, United States of America\\
% ...
$^{116}$ LAL, Universit{\'e} Paris-Sud and CNRS/IN2P3, Orsay, France\\
% ...
$^{159}$ Department of Physics, University of Toronto, Toronto ON, Canada\\
% ...
\end{flushleft}


\end{document}